\documentclass[letterpaper,12pt]{article}

\usepackage{graphicx}
\usepackage{grffile}
\usepackage{amsmath}
\usepackage{fancyhdr}
\usepackage{enumerate}
\usepackage{indentfirst}
\usepackage[hang,flushmargin]{footmisc} 
\usepackage{bibentry}
\usepackage{xspace}
\usepackage{geometry}
\usepackage[
  pdfusetitle,
  pdfkeywords={LAr1, neutrino, oscillation, sterile},
  bookmarks,bookmarksopen,
  bookmarksnumbered,
  pdfpagelabels,
  colorlinks,linkcolor=blue,citecolor=red,
  pdfview={Fit},
]
{hyperref}

\graphicspath{./}

\newcommand{\MB}{MiniBooNE\xspace}
\newcommand{\SB}{SciBooNE\xspace}
\newcommand{\harp}{HARP\xspace}

\newcommand{\uboone}{\text{MicroBooNE}\xspace}
\newcommand{\lartpc}{\text{LArTPC}\xspace}
\newcommand{\lartpcs}{\text{LArTPCs}\xspace}
\newcommand{\larnd}{\text{LAr1-ND}\xspace}
\newcommand{\larfd}{\text{LAr1-FD}\xspace}
\newcommand{\larone}{\text{LAr1}\xspace}

\newcommand{\rem}{\mathrm}

\newcommand{\ndpot}{\ensuremath{2.2 \times 10^{20}}\xspace}
\newcommand{\ubpot}{\ensuremath{6.6 \times 10^{20}}\xspace}
\newcommand{\dmsq}{\ensuremath{\Delta m^2}\xspace}

\newcommand{\sinth}{\ensuremath{\sin^22\theta}\xspace}

\newcommand{\numunue}{\ensuremath{\numu \rightarrow \nue}\xspace}

\newcommand{\numunuebar}{\ensuremath{\bar{\nu}_{\mu} \rightarrow \bar{\nu}_{e}}\xspace}

\newcommand{\numunux}{\ensuremath{\numu \rightarrow \nu_x}\xspace}
\newcommand{\chisq}{\ensuremath{\chi^{2}}\xspace}

\newcommand{\Enu}{\ensuremath{E_{\nu}}\xspace}

\newcommand{\lenu}{\ensuremath{L/E_{\nu}}\xspace}

\newcommand{\numu}{\ensuremath{\nu_{\mu}}\xspace}
\newcommand{\nue}{\ensuremath{\nu_{e}}\xspace}
\newcommand{\nutau}{\ensuremath{\nu_{\tau}}\xspace}
\newcommand{\numubar}{\ensuremath{\bar\nu_{\mu}}\xspace}
\newcommand{\nuebar}{\ensuremath{\bar\nu_{e}}\xspace}

\newcommand{\nuone}{\ensuremath{\nu_{1}}\xspace}
\newcommand{\nutwo}{\ensuremath{\nu_{2}}\xspace}
\newcommand{\nuthree}{\ensuremath{\nu_{3}}\xspace}
\newcommand{\nufour}{\ensuremath{\nu_{4}}\xspace}
\newcommand{\nufive}{\ensuremath{\nu_{5}}\xspace}

\newcommand{\pizero}{\ensuremath{\pi^{0}}\xspace}
\newcommand{\kplus}{\ensuremath{\mbox{K}^{+}}\xspace}

\newcommand{\kzero}{\ensuremath{\mbox{K}^{0}}\xspace}
\newcommand{\prot}{\ensuremath{\mbox{p}}\xspace}

\newcommand{\muminus}{\ensuremath{\mu^{-}}\xspace}

\newcommand{\GeV}{\ensuremath{\mbox{GeV}}\xspace}
\newcommand{\MeV}{\ensuremath{\mbox{MeV}}\xspace}
\newcommand{\GeVc}{\ensuremath{\mbox{GeV}/c}\xspace}

\newcommand{\eVsq}{\ensuremath{\mbox{eV}^2}\xspace}

\newcommand{\m}{\ensuremath{\mbox{m}}\xspace}

\usepackage{authblk}

\author[1]{C.~Adams}
\author[2]{C.~Andreopoulos}
\author[3]{J.~Asaadi}
\author[4]{B.~Baller}
\author[5]{M.~Bishai}
\author[6]{L.~Camilleri}
\author[1]{ F.~Cavanna}
\author[5]{H.~Chen}
\author[1]{E.~Church}
\author[7]{D.~Cianci}
\author[8]{G.~Collin}
\author[8]{J.~Conrad}
\author[9]{A.~Ereditato}
\author[1]{B.~Fleming\footnote{Contact persons}}
\author[7]{W.M.~Foreman}
\author[10]{G.~Garvey}
\author[11]{R.~Guenette}
\author[8]{C.~Ignarra}
\author[8]{B.~Jones}
\author[6]{G.~Karagiorgi}
\author[10]{W.~Ketchum}
\author[9]{I.~Kreslo}
\author[5]{D.~Lissauer}
\author[10]{W.C.~Louis}
\author[2]{K.~Mavrokoridis}
\author[2]{N.~McCauley}
\author[10]{G.B.~Mills}
\author[1]{ O.~Palamara*}
\author[10]{Z.~Pavlovic}
\author[5]{X.~Qian}
\author[10]{L.~Qiuguang}
\author[4]{R.~Rameika}
\author[7]{D.W.~Schmitz*}
\author[6]{M.~Shaevitz} 
\author[3]{M.~Soderberg}
\author[8]{J.~Spitz}
\author[1]{A.M.~Szelc}
\author[10]{C.E.~Taylor}
\author[6]{K.~Terao}
\author[12]{M.~Thomson}
\author[5]{C.~Thorn}
\author[8]{M.~Toups}
\author[2]{C.~Touramanis}
\author[9]{T.~Strauss}
\author[10]{R.G.~Van~De~Water}
\author[9]{C.R.~von~Rohr}
\author[9]{M.~Weber}
\author[5]{B.~Yu}
\author[4]{G.~Zeller}
\author[7]{J.~Zennamo}

\affil[1]{\em\small Yale University, New Haven, CT}
\affil[2]{\em University of Liverpool, Liverpool, UK}
\affil[3]{\em Syracuse University, Syracuse, NY}
\affil[4]{\em Fermi National Accelerator Laboratory, Batavia, IL}
\affil[5]{\em Brookhaven National Laboratory, Upton, NY}
\affil[6]{\em Columbia University, Nevis Labs, Irvington, NY}
\affil[7]{\em University of Chicago, Enrico Fermi Institute, Chicago, IL}
\affil[8]{\em Massachusetts Institute of Technology, Boston, MA}
\affil[9]{\em Laboratory for High Energy Physics, University of Bern, Switzerland}
\affil[10]{\em Los Alamos National Laboratory, Los Alamos, NM}
\affil[11]{\em University of Oxford, Oxford, UK}
\affil[12]{\em University of Cambridge, Cambridge, UK}

\begin{document}

\pagestyle{empty}



\title{\sf \larnd: Testing Neutrino Anomalies with \\ \vspace{1ex}
Multiple \lartpc Detectors at Fermilab}
\maketitle



\tableofcontents
\clearpage

\setcounter{page}{1}
\pagestyle{fancy}
\rhead{\larnd White Paper / \thepage}
\lhead{}
\cfoot{}

\vspace*{-5ex}

\section{Introduction and Overview}
\label{sec:intro}


Neutrino oscillations, for which convincing evidence has emerged only in the past fifteen years \cite{nuosc}, stand as one of the most important recent discoveries in particle physics.  Through the phenomenon of oscillations, we are now presented with exciting new opportunities for continued discovery. 
First, large mixing between the three active flavors is now confirmed \cite{theta13}.  This opens the door to a new era of discovery in long-baseline experimental neutrino physics, including the opportunity to test for CP-violation in the neutrino sector \cite{lbne}.  Testing for CP-violation among the neutrinos is an important step in determining if leptogenesis \cite{ref:lepto} is a viable explanation for the matter--antimatter asymmetry that clearly exists in the universe today.  Second, there are existing experimental anomalies at short-baselines that may be hinting at exciting new physics, including the possibility of additional low-mass sterile neutrino states.   Definitive evidence for sterile neutrinos would be a revolutionary discovery, with implications for particle physics as well as cosmology.   The recent anomalous experimental results, briefly reviewed below, have piqued interest within the community for pursuing experiments that can provide new input to this important puzzle \cite{sterile}.  

Proposals to address these hints from reactor, accelerator, and source experiments are in the planning stages or underway worldwide.  With the MicroBooNE experiment, currently under construction on the Booster Neutrino Beamline, Fermilab will be the first to explore these hints in neutrinos.  This proposal aims to build upon that program and significantly extend the physics reach of the Booster Neutrino Beam facility by adding a modest-scale \lartpc near detector, \larnd.  

\larnd can serve as the next step in a phased program to build a world-class short-baseline neutrino program at Fermilab.  The combination of \larnd and \uboone will provide a clear interpretation of the anomalous signal observed by \MB in neutrinos and enable important additional opportunities to test for new physics through disappearance channels.  The full LAr1 experiment, previously presented in an LOI \cite{lar1loi} and a Snowmass Neutrino Group white paper \cite{lar1white}, can definitively address the sterile neutrino question in both neutrinos and antineutrinos.  In addition, starting with the phase-1 \larnd detector the LAr1 program serves as a valuable development project for LArTPC technology toward future experiments.

\subsection{Liquid Argon TPCs for Neutrino Physics}

Liquid argon time project chamber detectors (\lartpcs) are particularly attractive for use in neutrino physics because of their exceptional capabilities in tracking, particle identification and calorimetric energy reconstruction.  The idea of using such a device for neutrino detection was first proposed by Carlo Rubbia over 35 years ago \cite{rubbia}. 
Charged particles propagating in liquid argon will ionize nearby atoms and the freed electrons drift under the influence of an electric field ($\sim$500 V/cm) applied across the detector volume.  At the detector boundary planes of closely spaced sense wires (wire pitch in the 3-5 mm range) are used to collect the free charge. 
Signals read out on the wires are proportional to the amount of energy deposited in that spatial region. 

\begin{figure}[t]
\begin{center}
\mbox{ \includegraphics[width=0.5\textwidth]{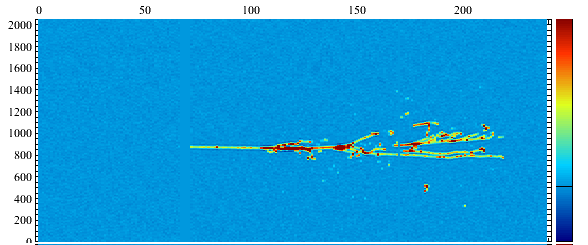} 
\includegraphics[width=0.5\textwidth]{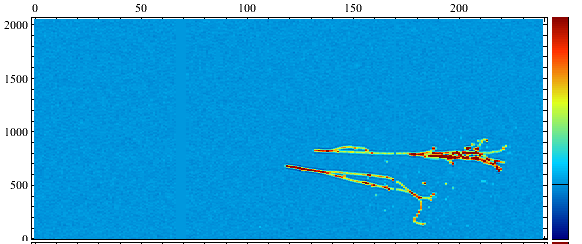}}
\caption{\small Examples of simulated particle interactions in a \lartpc detector illustrating the power to dustinguish electrons from photons.  The left panel shows a single 1 GeV electron shower.  The right panel shows the decay of a 1 GeV \pizero to two photons.  The horizontal axis is the channel number on the collection wire plane.  The vertical axis is the hit time and the color indicates the amplitude of the signal on that wire on a linear scale such that dark red is approximately double the energy deposit of the green.}
\label{fig:lartpc}
\end{center}
\end{figure}

The fine sampling and calorimetry enabled by this technology is key to its performance in neutrino physics.  Cherenkov detectors, such as \MB and T2K, are not able to distinguish electrons from single photons.  An important advantage of the LArTPC is the ability to separate electrons and photons by sampling the energy deposition before the buildup of an electromagnetic shower.  When a photon converts to an $e^{+}e^{-}$ pair, the resulting ionization in the first few centimeters is consistent with two minimum ionizing particles (mips), distinguishing it from the single mip deposit of electrons.  Figure \ref{fig:lartpc} compares a simulated single electron to a \pizero which has decayed to two photons. \pizero decays are an important source of backgrounds in searches for electron neutrinos, especially when one of the photons does not convert in the detector.  In the right figure, both photons from a \pizero decay are clearly visible in the event display, but, more importantly, the ionization strength near the beginning of each shower is approximately double that of the same portion of the electron shower on the left.  This double mip deposition of converted photons allows even single photons to be identified, significantly reducing backgrounds in searches for \nue appearance relative to other technologies.   

\subsection{A Staged Multi-LArTPC Program at Fermilab}
\label{sec:staged}

The use of multiple detectors at different baselines presents a significant advantage for reducing systematic uncertainties in the measurement of neutrino oscillations; MINOS and Daya Bay are recent examples of the power of this approach \cite{theta13}.
The anomalous short-baseline results discussed here may be hinting at neutrinos oscillating with an amplitude 10 to 100 times smaller than those measured at Daya Bay and MINOS, thus emphasizing further the need for a multiple detector experiment in conducting a sensitive search for sterile neutrinos.



Fermilab has an opportunity to pursue such a multi-detector program using the superior event reconstruction capabilities of the \lartpc technology through the construction of new detectors in the existing Booster Neutrino Beam.
In a first phase, 
a modest-scale \lartpc near detector, \larnd, 
can determine the nature of an excess in neutrino mode, if observed by MicroBooNE.  If not observed, this near detector can, in combination with MicroBooNE, rule out the low-energy excess observed by \MB \cite{mbosc1, mbosc2} as a hint for short-baseline neutrino oscillations and fill in a crucial piece of the puzzle created by the existing anomalies.  
A near detector also extends the physics program in the Booster Beam by making possible a sensitive test of \numu disappearance through charged-current interactions, as well as a search for active flavor disappearance through neutral-current channels. These are critical aspects of a search for oscillations to sterile neutrinos and are only enabled with a near detector.

\begin{figure}[t]
\begin{center}
\includegraphics[width=0.9\textwidth]{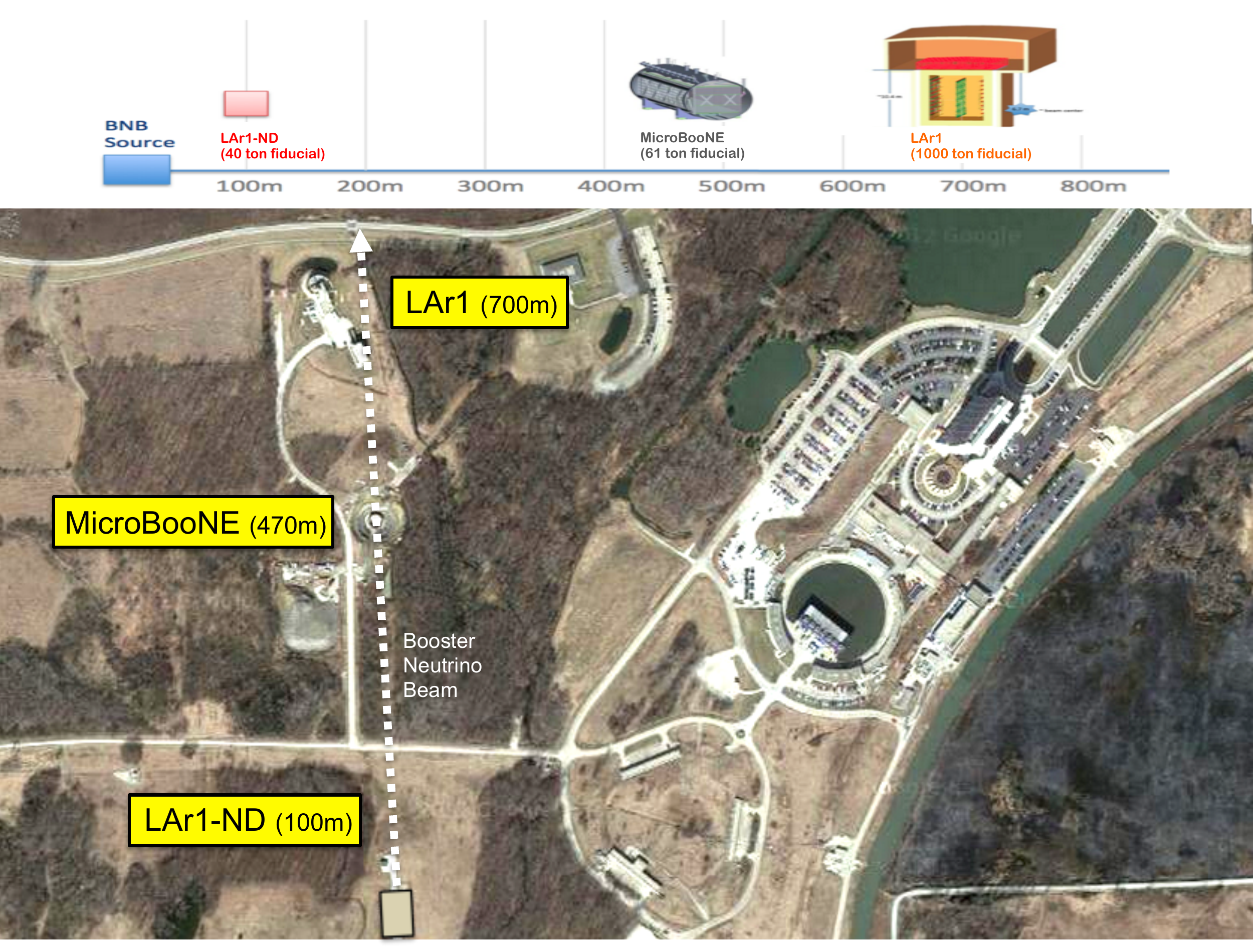}
\caption{\small Aerial view of the Fermilab Booster Neutrino Beam showing the locations of the \uboone detector (currently under construction), the proposed \larnd, and a possible future kiloton-scale \lartpc detector, LAr1.}
\label{fig:baseline}
\end{center}
\end{figure}

LAr1-ND is a $\sim$40 ton fiducial volume LArTPC to be located in the existing \SB enclosure.  
The cryostat, cryogenics system, TPC and light collection systems, high voltage configuration and electronics readout will all serve as development steps towards LBNE.    The membrane cryostat will house CPAs (Cathode Plane Assemblies) and APAs (Anode Plane Assemblies) to read out ionization electron signals.  The front end electronics will build on the MicroBooNE and LBNE designs with cold pre-amplifiers multiplexed within LAr.  The cryogenics system and high voltage configuration will take advantage of that learned from LAPD (the Liquid Argon Purity Demonstrator), the 35 ton membrane cryostat prototype, and MicroBooNE.  Overall, the design philosophy of the LAr1-ND detector is as a prototype for LBNE that functions as a physics experiment.  While the present conceptual design described here is an excellent test of LBNE detector systems sited in a neutrino beam, the \larnd collaboration is exploring innovations in this design and the opportunity to test them in a running experiment.  LAr1-ND is an opportunity, therefore, to further the development of the LBNE detector design. 

Beyond this near-term program, LAr1-ND can serve as a near detector for the full LAr1 program to definitively address the MiniBooNE antineutrino anomaly \cite{mbosc3, mbcomb} and make precision measurements of high-\dmsq neutrino oscillations.  

In summer 2012, the LAr1 collaboration submitted a Letter of Intent \cite{lar1loi} to the Fermilab PAC describing the physics reach of a 1 kton LArTPC detector located at 700 m along the Booster Neutrino Beam to serve as a far detector for the MicroBooNE experiment.  Figure \ref{fig:baseline} shows the locations of the \uboone detector, the LAr1 detector and the \larnd described here.  This multiple detector combination would provide a powerful constraint of experimental uncertainties and address in a definitive way whether the observed short-baseline anomalies are due to high \dmsq neutrino oscillations.  

Since the LAr1 LOI, the advantages of starting with LAr1-ND as a first phase both for timeliness and efficiency have become clear.  LAr1-ND, in combination with \uboone, will bring a compelling early physics program, decisive determination of the nature of the neutrino anomaly, and serve as a development step towards LBNE both for hardware and software.  A new project at this scale also provides an avenue for expanding expertise with \lartpc detectors within the neutrino physics community and an opportunity for building international collaboration in U.S. experimental neutrino physics.  



With the existing SciBooNE enclosure at the appropriate near location for the LAr1-ND program, advanced design work from both MicroBooNE and LBNE, and relatively modest cost, the LAr1-ND can be built quickly and given the envisioned fiducial volume, have a definitive result within one year.  With this schedule, the data run can be concurrent with the final year of MicroBooNE data taking, opportunistically taking advantage of the already approved MicroBooNE run.  This timeliness puts Fermilab in an excellent position to not only confirm or rule out MiniBooNE's neutrino anomaly, but be able to interpret it as an oscillation signal.  The full LAr1 program can then proceed to address the antineutrino results.

\section{Motivation: Short-Baseline Anomalies in Neutrino Physics}
\label{sec:Motivation}

In this section we very briefly review the various experimental anomalies that hint at the possibility of new physics occurring in the neutrino sector.  We include this description for completeness, but refer the reader to more thorough descriptions of these results and their interpretation, including \cite{sterile, kopp} and the references therein. 

In recent years, experimental anomalies ranging in significance (2.8--3.8$\sigma$) have been reported from a variety of experiments studying neutrinos over baselines less than 1~km.  Results from the LSND and MiniBooNE short-baseline \nue/\nuebar appearance experiments 
show anomalies which cannot be described by oscillations between the three standard model neutrinos (the ``LSND anomaly'').  In addition, a re-analysis of the antineutrino flux produced by nuclear power reactors has led to an apparent deficit in \nuebar event rates in a number of reactor experiments (the``reactor anomaly'').   Similarly, calibration runs using $^{51}$Cr and $^{37}$Ar radioactive sources in the Gallium solar neutrino experiments GALLEX and SAGE have shown an unexplained deficit in the electron neutrino event rate over very short distances (the ``Gallium anomaly'').  




\subsection*{LSND \numunuebar}

The Liquid Scintillator Neutrino Detector experiment (LSND) was conducted at Los Alamos
National Laboratory from 1993 through 1998.  LSND used a decay-at-rest (DAR) pion beam to produce a beam of \numubar between 20-53~MeV about 30~m from a liquid scintillator-based detector.  \nuebar were detected through inverse beta decay (IBD) on carbon, $\nuebar p \rightarrow e^{+}n$.
The signature of IBD events is a prompt positron followed by a 2.2 MeV $\gamma$ produced when the neutron captures on free protons in the scintillator.  After 5 years of data taking, 89.7$\pm$ 22.4 $\pm 6.0$ $\overline{\nu}_e$  candidate events were observed above backgrounds, corresponding to 3.8$\sigma$ evidence for $\overline{\nu}_{\mu} \rightarrow \overline{\nu}_{e}$ oscillations  \cite{lsnd} occurring at a \dmsq in the 1 \eVsq region.  This signal, therefore, cannot be accommodated within the three Standard Model neutrinos, and like the other short baseline hints for oscillations at \lenu $\sim $1~m/\MeV, implies new physics.

\begin{figure}
\centering
\mbox{ \includegraphics[width=0.37\textwidth, trim=1cm -1.5cm 0cm 0cm]{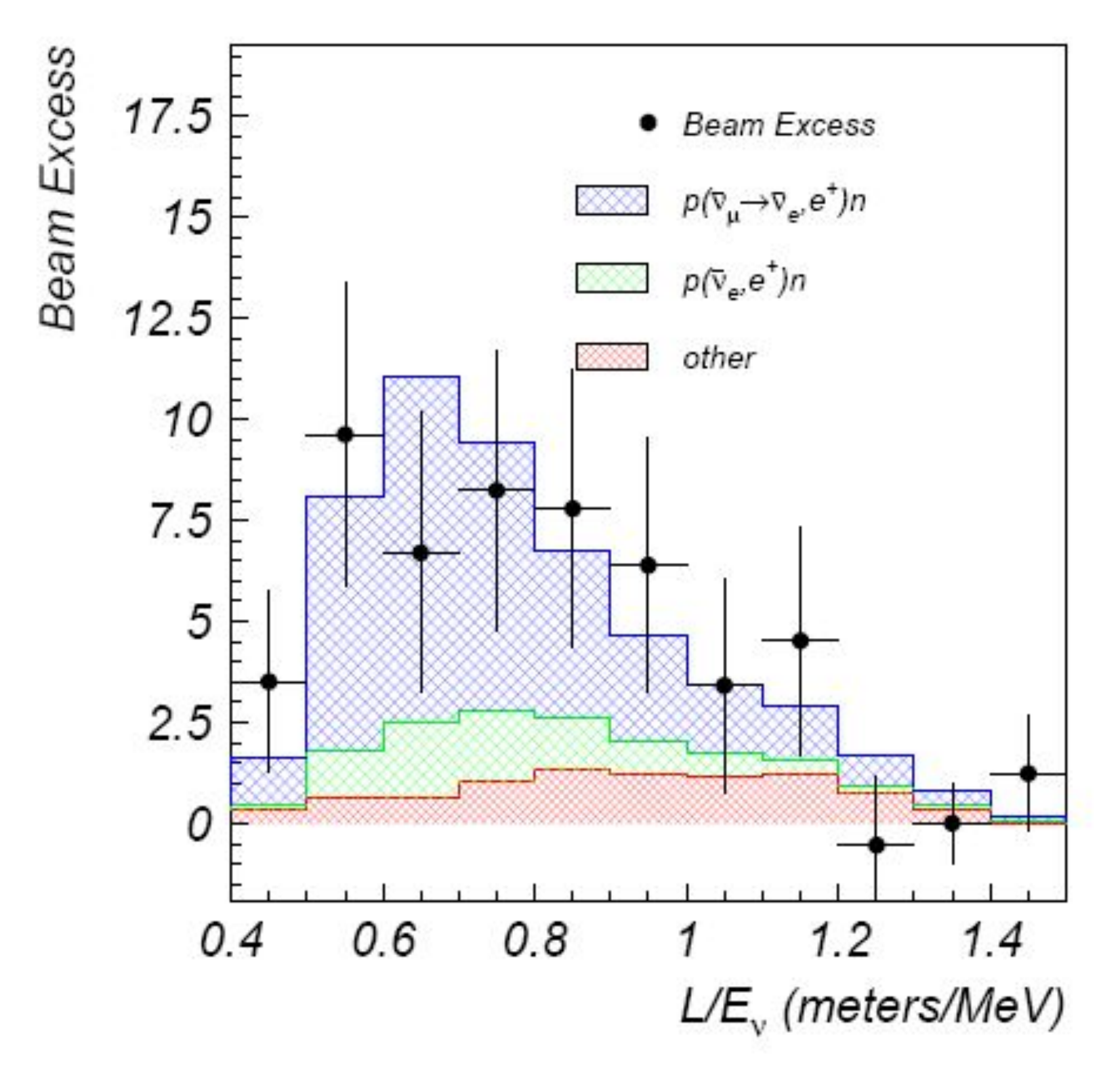} \quad
\includegraphics[width=0.6\textwidth, trim=1cm 1.2cm 1cm 1cm]{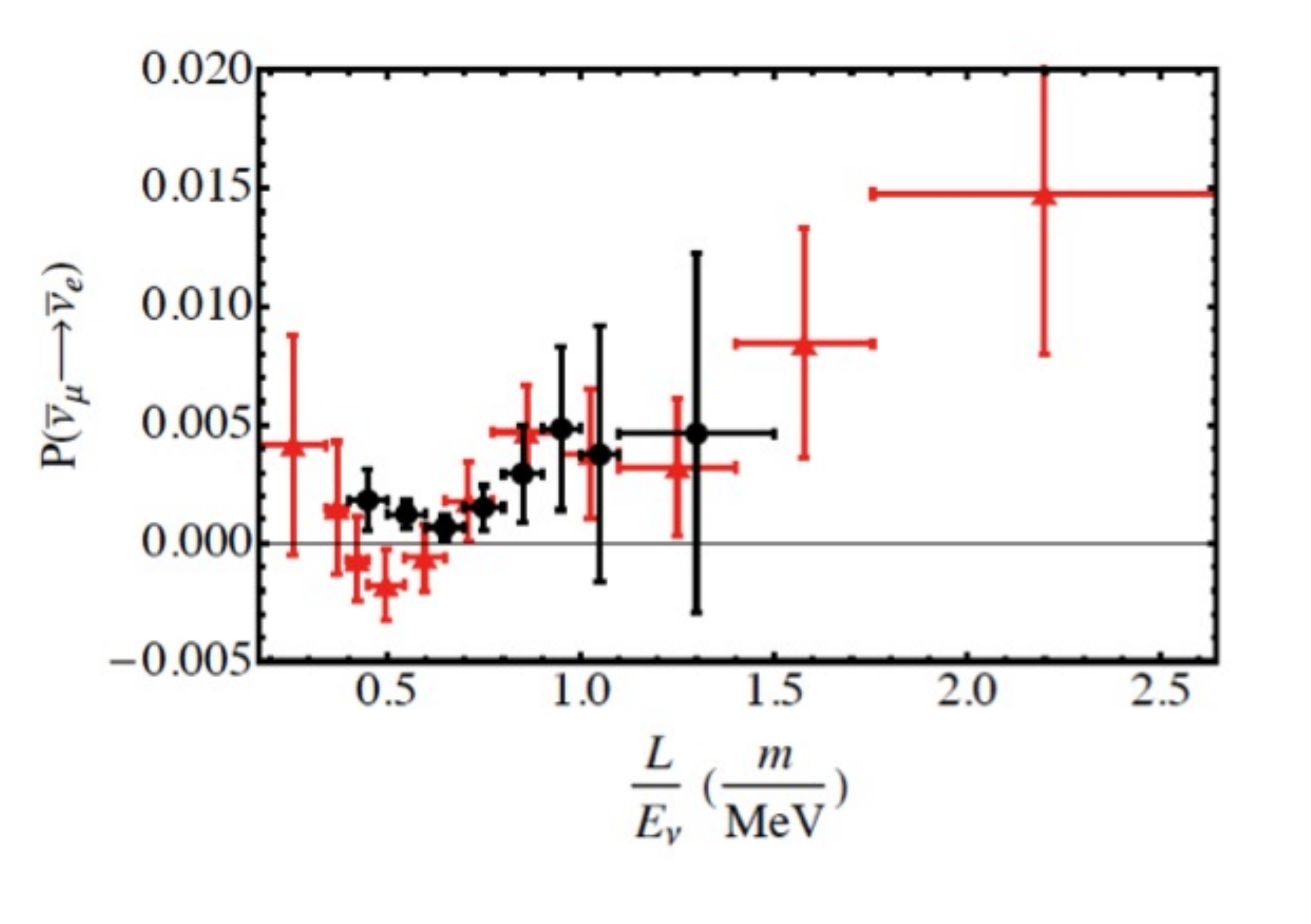}}
\caption{\small Left: Excess of electron neutrino candidate events observed by the LSND experiment.  Right: Oscillation probability as a function of \lenu if the excess candidate events are assumed due to \numunuebar transitions using MiniBooNE (red) and LSND (black) data. }
\label{fig:mbooneLSND}
\end{figure}

\subsection*{MiniBooNE \numunue and \numunuebar}

\begin{figure}[tb]
\begin{center}
\mbox{ \includegraphics[height=0.24\textheight, trim=10mm 0mm 10mm 11mm, clip]{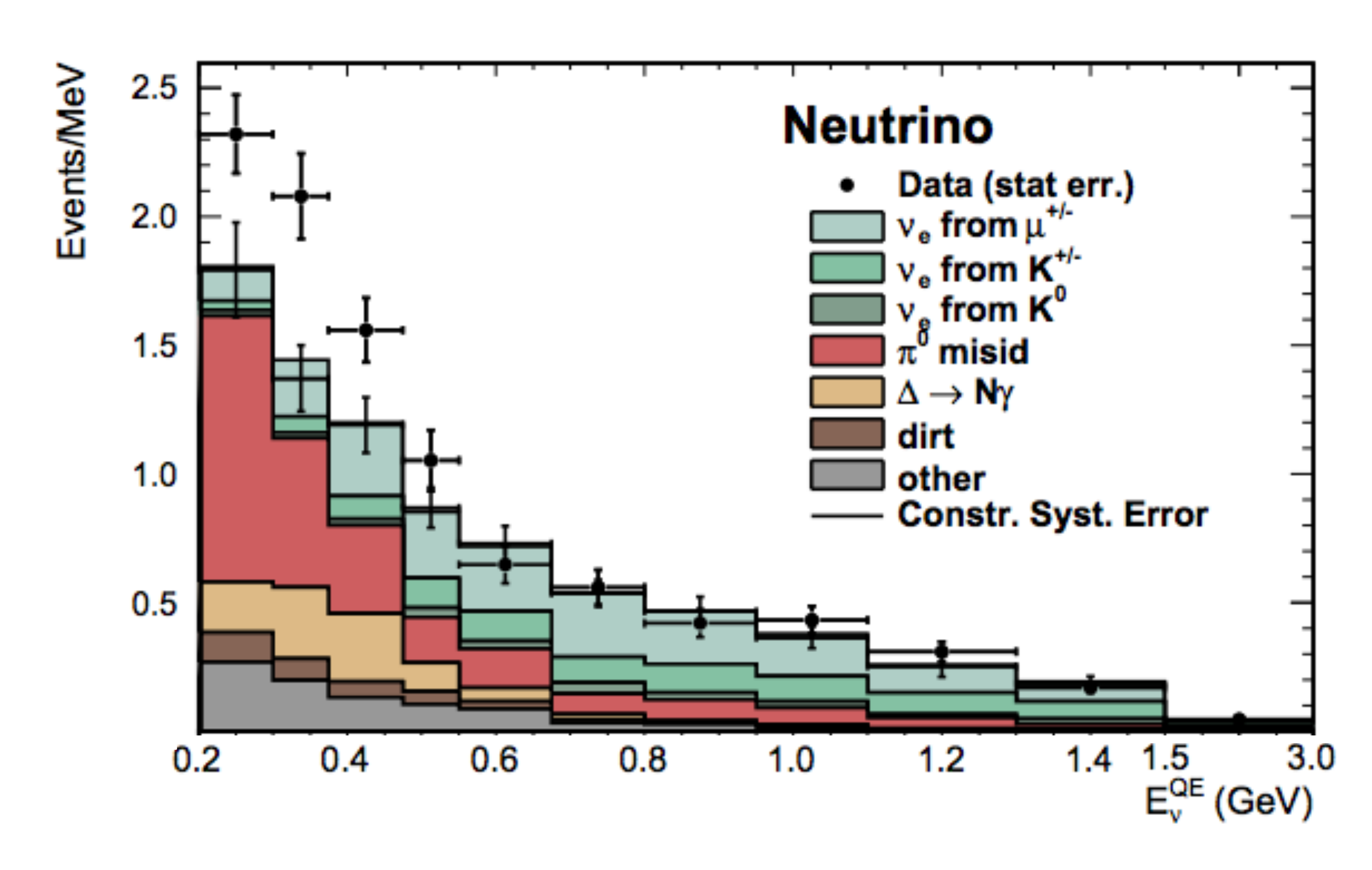} 
\includegraphics[height=0.24\textheight, trim=0mm 0mm 10mm 0mm, clip]{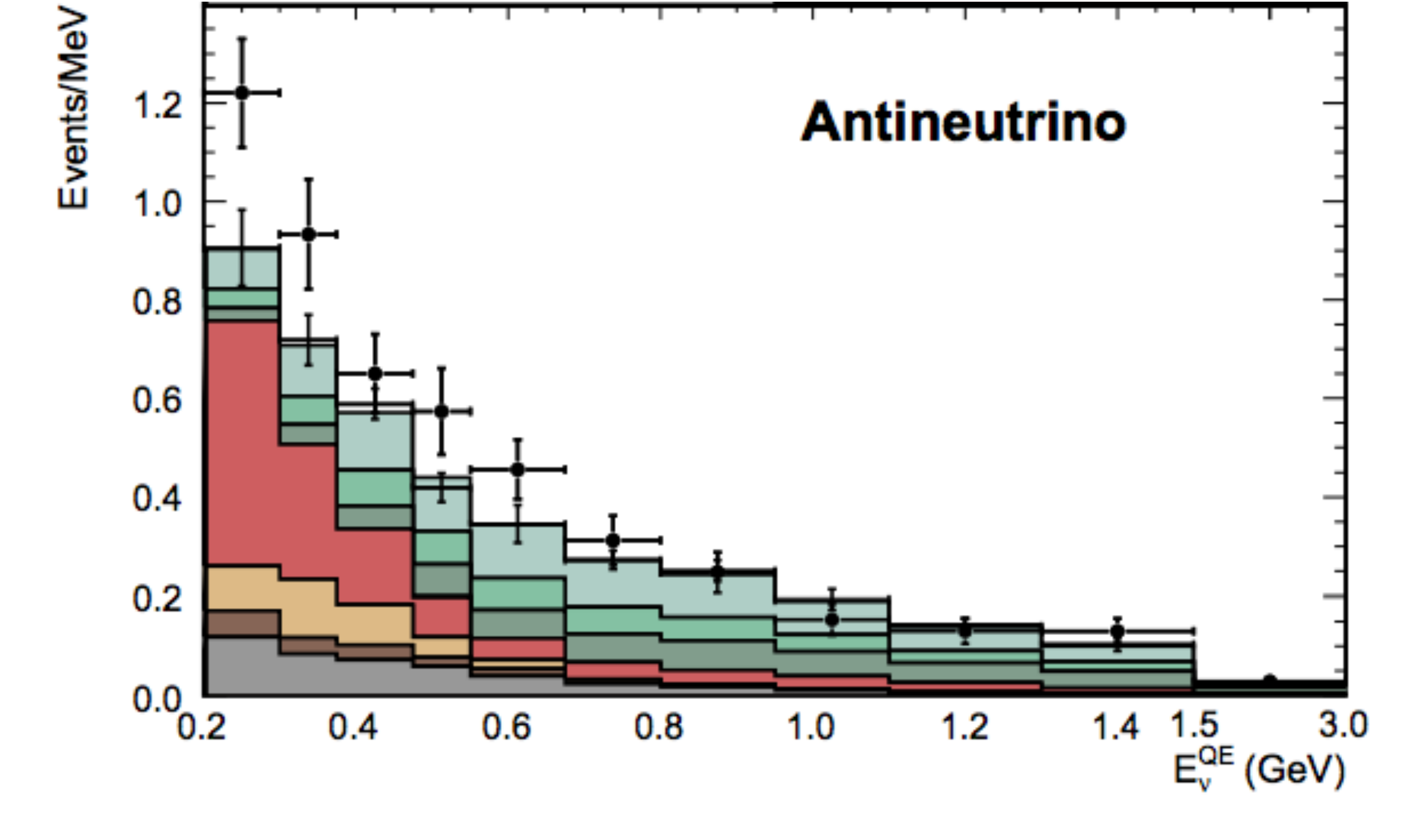}}
  \caption{\small The neutrino mode (left) and antineutrino mode (right) electron neutrino candidate distributions from \MB.  The stacked histograms show the background expectation with systematic errors and the black points are the data with statistical errors.  Plots taken from \cite{mbcomb}.}
  \label{fig:mboone}
  \end{center}
\end{figure}

The MiniBooNE collaboration has recently completed an analysis of their full ten year data set including both neutrino and antineutrino running \cite{mbcomb}.   The \MB detector sits $\sim$500 m downstream of the Booster Neutrino Beam at Fermilab.  Predominately muon flavor neutrinos are produced in pion decay-in-flight, yielding a broad beam with peak energy around 700~\MeV.  Muon and electron neutrinos are identified in charged-current interactions by the characteristic signatures of Cherenkov rings for muons and electrons.  

MiniBooNE observes a 3.4$\sigma$ signal excess of \nue candidates at low energy in neutrino mode (162.0 $\pm$ 47.8 electromagnetic events at reconstructed $E_\nu$ below 475~MeV).  These events, along with backgrounds, are shown in Figure~\ref{fig:mboone}.   The excess events can be electrons or single photons since these are indistinguishable in MiniBooNE's Cherenkov imaging detector.  \uboone will address this question at the same baseline as \MB by applying the \lartpc technology to separate electrons and gammas.  \larnd at 100~m, in combination with \uboone, enables a powerful test of the nature of any excess found by determining if it exists intrinsically in the Booster Neutrino Beam.  

MiniBooNE also observes an excess of 78.4 $\pm$ 28.5 electron antineutrino candidates (2.8$\sigma$) at both low and somewhat higher energies than in neutrinos as shown in Figure~\ref{fig:mboone}.  Figure~\ref{fig:mbooneLSND} compares the $L/E_{\nu}$ dependence of these events to the excess observed at LSND.  It is this signal in antineutrino mode that the full multi-detector LAr1 experiment could address.


\subsection*{Reactor neutrino anomaly}

\begin{figure}
\centering
\includegraphics[width=1.0\textwidth]{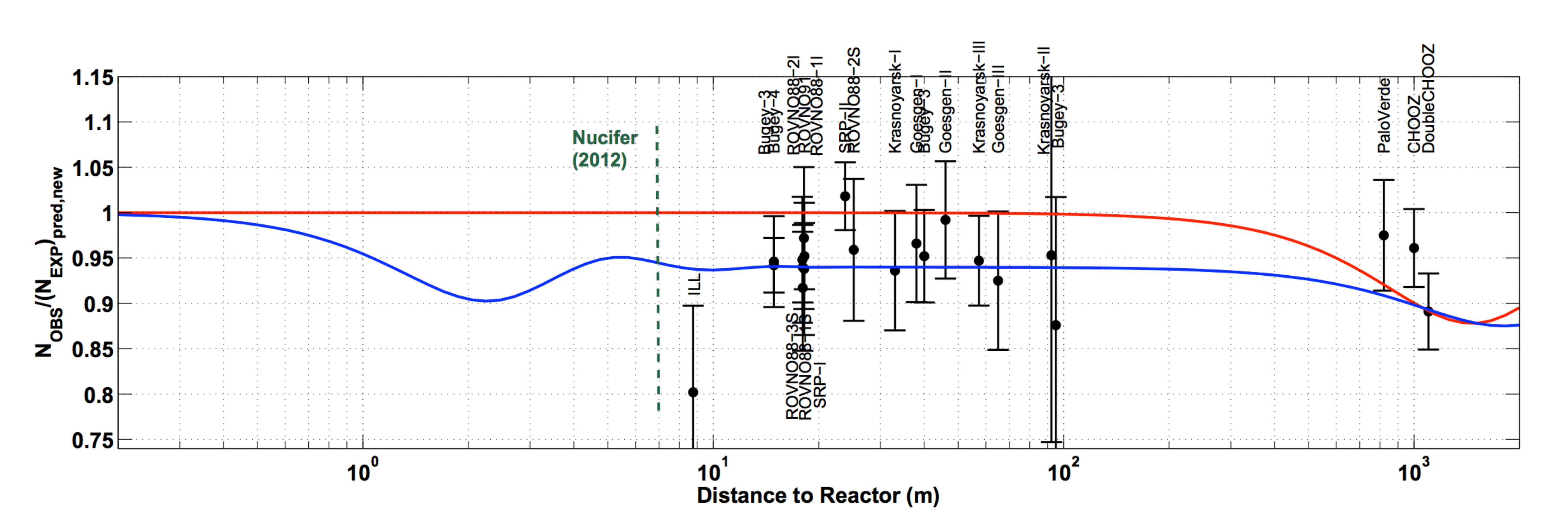}
\caption{\small Ratio of the observed to predicted reactor \nuebar rate for 19 different reactor neutrino experiments at baselines less that 100~m.  The mean average ratio including correlations is $0.927\pm0.023$, indicating a $\sim7\%$ deficit at short baseline.  
The curves show fits to the data assuming standard three neutrino oscillations (red) and assuming 3+1 neutrino oscillations including one additional sterile neutrino (blue)~\cite{reactor}.
}
\label{fig:reactors}
\end{figure}

A re-evaluation of the \nuebar flux produced by nuclear power reactors \cite{reactorspectra1, reactorspectra2} has prompted a re-analysis of short baseline reactor $\overline{\nu}_e$ disappearance measurements from the last several decades~\cite{reactor}.  
The new reference spectra takes advantage of a re-evaluation of inverse beta decay cross sections impacting the neutron lifetime, and accounts for long-lived radioisotopes accumulating in reactors.  Figure~\ref{fig:reactors} shows this predicted flux compared to reactor measurements as a function of the baseline of each experiment. With this new prediction, the observed rates of interactions in detectors between 10 and 100 meters from the reactors are, on average, 6-7\% lower than that expected in the absence of oscillations.
This result can be explained through \nuebar disappearance due to oscillations at \dmsq $\sim$1 \eVsq, which could be consistent with the \MB and LSND appearance anomalies.    


\subsection*{GALLEX and SAGE calibration data}

\begin{figure}[h]
\begin{center}
\begin{tabular}{cc}
\includegraphics[angle=0,width=0.6\textwidth]{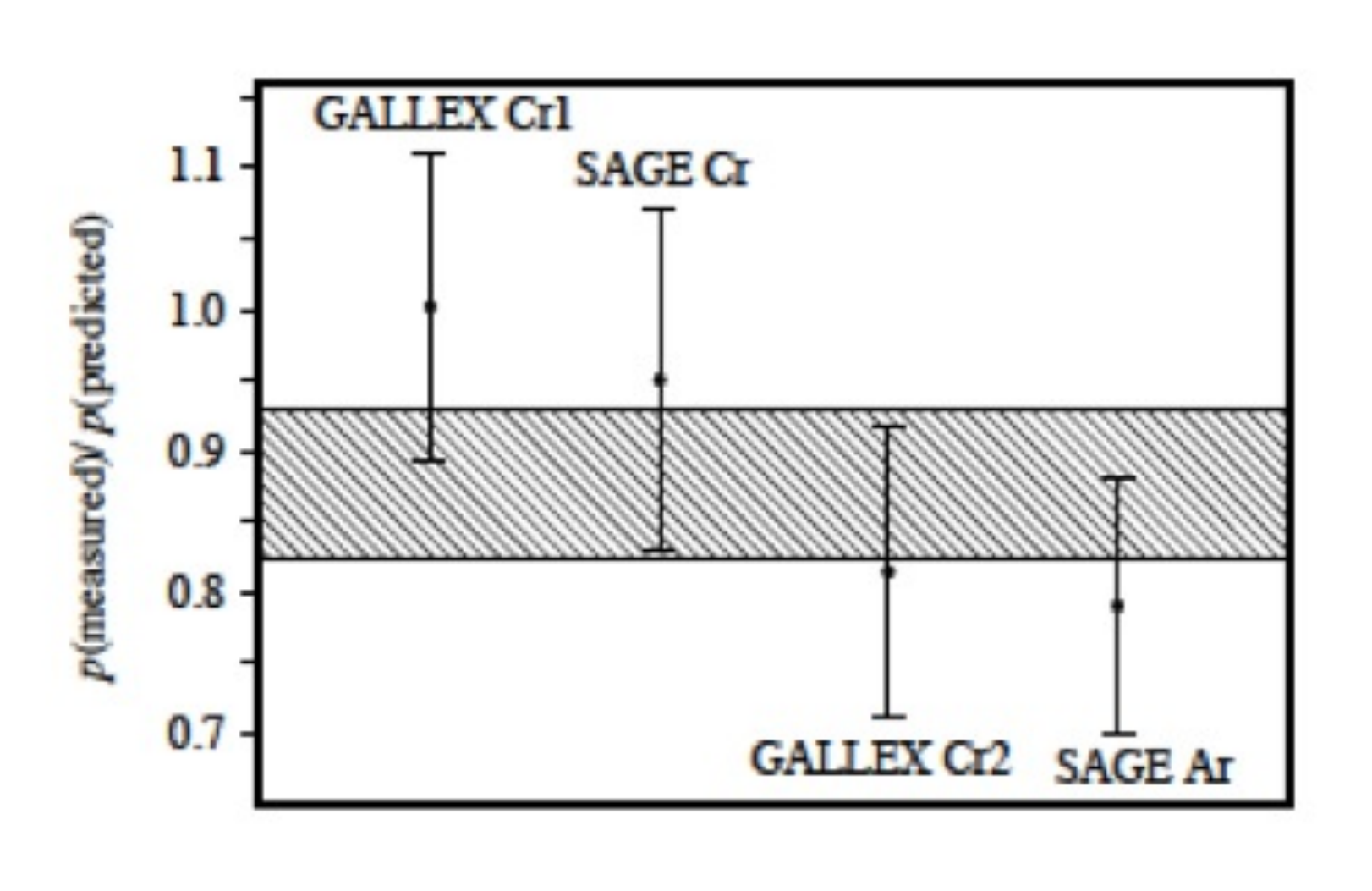}
\end{tabular}
\caption{\small The measured/predicted event ratio for GALLEX and SAGE source calibration data.  The average, shown by the shaded band, is 0.86 $\pm$ 0.05. }
\label{fig:GallexSage}
\end{center}
\end{figure}

Both the GALLEX and SAGE solar neutrino experiments used test sources  to calibrate their detectors.  In total they ran 4 test runs, 2 in GALLEX and 1 in SAGE with a $^{51}$Cr source which emits a 750~keV \nue, and 1 in SAGE with a $^{37}$Ar source, an 810~keV \nue emitter.  The test data reveal a deficit of electron neutrinos relative to the predicted rate as shown in  Figure~\ref{fig:GallexSage}.  The best fit ratio of data to prediction is 0.86 $\pm$ 0.05~\cite{Gallex, Sage}.  This deficit of very low energy electron neutrinos over very short baselines could also be explained through \nue disappearance due to oscillations at \dmsq $\geq$ 1 \eVsq.


\subsection*{Interpretation}

Table~\ref{summaryresults} summarizes the results discussed above and lists their individual significance. 
While each of these measurements taken separately lack the significance to claim a discovery, together these signals could be hinting at important new physics.  The most common interpretation is as evidence for the existence of one or more additional, mostly "sterile" neutrino states with masses at or below the few eV range.  In these models, the mass states \nuone, \nutwo and \nuthree are those responsible for the well established oscillations observed at $\dmsq_{21} = 7.5 \times10^{-5}$ \eVsq and  $\dmsq_{31} = 2.4 \times10^{-3}$ \eVsq and are taken to be dominated by active flavors (\nue, \numu, \nutau) with only small contributions from sterile flavors.  Additional higher mass neutrino states, \nufour, \nufive, ... are taken as mostly sterile with small active flavor content.  The experimental results described above can be interpreted as indications of oscillations due to mass-squared splitting in the $\dmsq_{41} \approx [0.1-10]~\eVsq$ range. 

It is clear that the existing anomalies must be explored further, both by repeating the existing measurements in an effective way, and by testing these results in new ways capable of addressing the oscillation hypothesis. 

\begin{table}[t]
\label{summaryresults}
\centering
\begin{tabular}{|c|c|c|c|}
\hline
Experiment	& Type	& Channel 	& Significance \\
\hline \hline
LSND &   DAR  &  \numunuebar CC  & 3.8$\sigma$   \\ \hline
MiniBooNE &   SBL accelerator   &  \numunue CC  & 3.4$\sigma$   \\ \hline
MiniBooNE &   SBL accelerator   &  \numunuebar CC  & 2.8$\sigma$   \\ \hline
GALLEX/SAGE &   Source - e capture  & \nue disappearance  &   2.8$\sigma$      \\ \hline
Reactors &   Beta-decay & \nuebar disappearance &  3.0$\sigma$     \\ \hline
\end{tabular}
\caption{\small Summary of the experimental hints suggesting the possibility of high-\dmsq neutrino oscillations.}
\end{table}





\section{\larnd Physics Program}

Full interpretation of any anomalous excess observed by MicroBooNE, regardless of whether it is electrons or photons, will require a second detector at a near location equally sensitive to the same interactions.  MicroBooNE alone can confirm the nature of the excess at the $\sim$500 m distance, but in either scenario the question of whether that excess appears over a distance or is intrinsic to the beam will need to be answered.  \larnd can address this important issue.  A near detector also extends the physics program in the Booster Beam by enabling a sensitive test of \numu disappearance through charged-current interactions as well as a search for active flavor disappearance through neutral-current channels.  These are critical aspects of a search for oscillations to sterile neutrinos and are only possible with a near detector.

In this section we explore the potential of augmenting the existing \uboone detector at 470~m with a smaller \lartpc near detector housed in the existing (but empty) SciBooNE detector hall located at 100~m.  
This is the experiment we are proposing here.  We present the sensitivity of several complimentary approaches to testing existing anomalies or searching for evidence of high-\dmsq neutrino oscillations.   

Section \ref{sec:flux} presents the neutrino fluxes at each detector location since understanding differences in flux shape is important when making comparisons between near and far detectors.  Section \ref{sec:sim} briefly describes the Monte Carlo used to perform the sensitivity studies.  In Sec. \ref{sec:lee} we present the ability to confirm an excess of electron neutrinos that matches exactly the anomalous excess reported by \MB. In Sec. \ref{sec:nueapp} we focus on testing a \numunue appearance scenario in the context of a 3+1 sterile neutrino model as was reported by LSND.  Section \ref{sec:numudis} shows the sensitivity reach of a two detector configuration to searching for \numu disappearance, a far less model-dependent test of \numunux oscillations.  A similarly important test of active to sterile oscillations can be performed using neutral-current interactions for which the sensitivity is described in Sec. \ref{sec:ncdis}.  In Sec. \ref{sec:photon} we discuss the ability to measure the cross section for single photon production should the low-energy excess be found to be photons instead of electrons.  Finally, Sec. \ref{sec:xsec} gives the expected event rates for different types of interactions, indicating the ability to make precise neutrino-argon cross section measurements in \larnd.  Later, in Sec. \ref{sec:PhaseII}, we briefly outline the physics reach of a possible future extension to this experimental program.  The addition of a 1~kton-scale far detector located at 700~m, \larfd, would provide a powerful opportunity to understand the antineutrino mode anomalies and potentially make precision measurements of oscillations to sterile states in neutrino mode.  

\subsection{Neutrino Fluxes}
\label{sec:flux}

The primary source of neutrinos for \larnd is the Booster Neutrino Beam (BNB). The same beamline has been used by the \MB experiment and will be used by \uboone.  In addition to the on-axis BNB flux, all detectors are also exposed to an off-axis component of the Neutrinos at the Main Injector (NuMI) beamline. While data from this off-axis beam will be studied and can improve the sensitivity to this physics, here we restrict our discussion to the on-axis flux from the BNB.

\begin{figure}[t]
\centering
\mbox{ \includegraphics[width=0.5\textwidth]{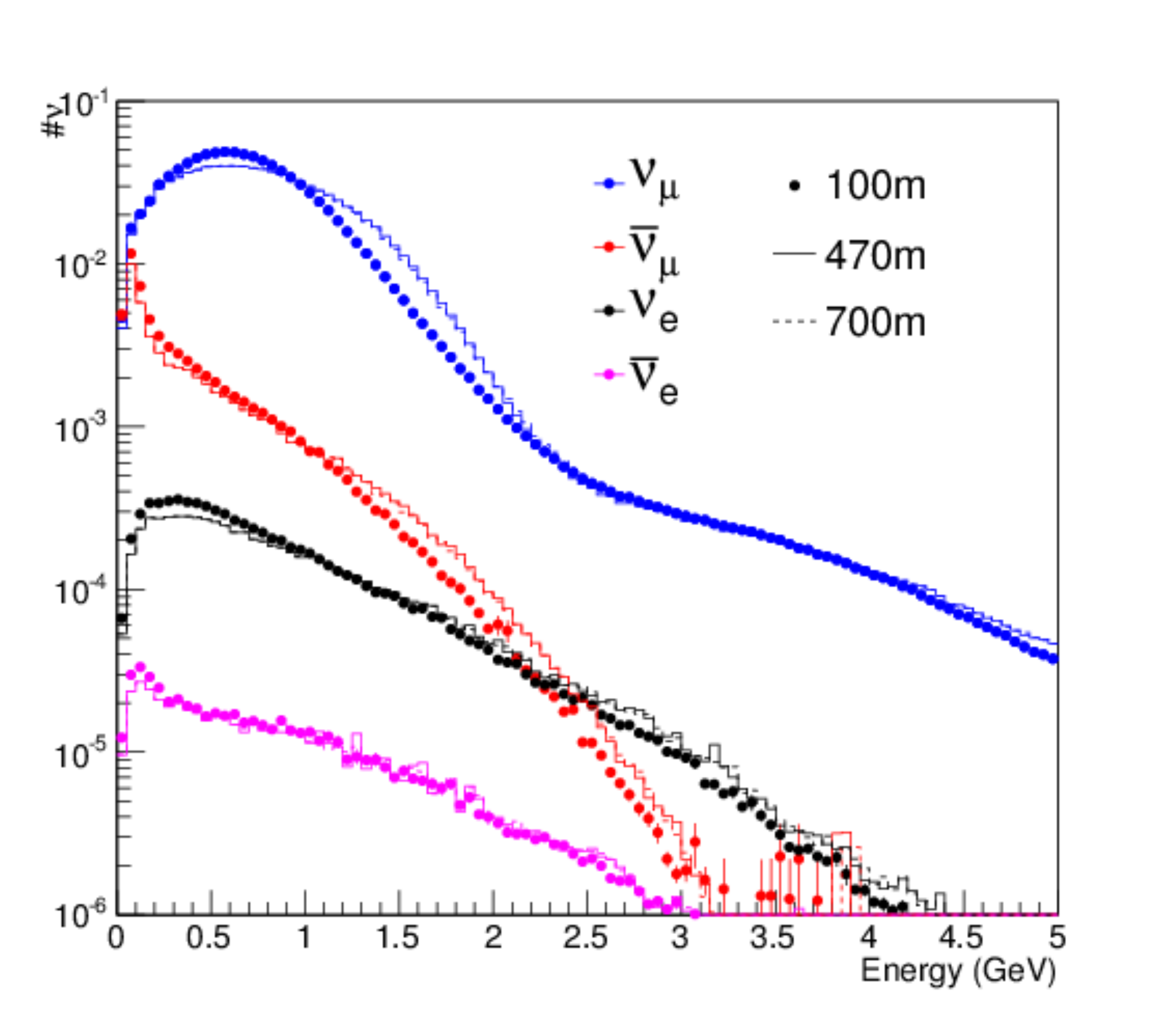} \quad
\includegraphics[width=0.5\textwidth]{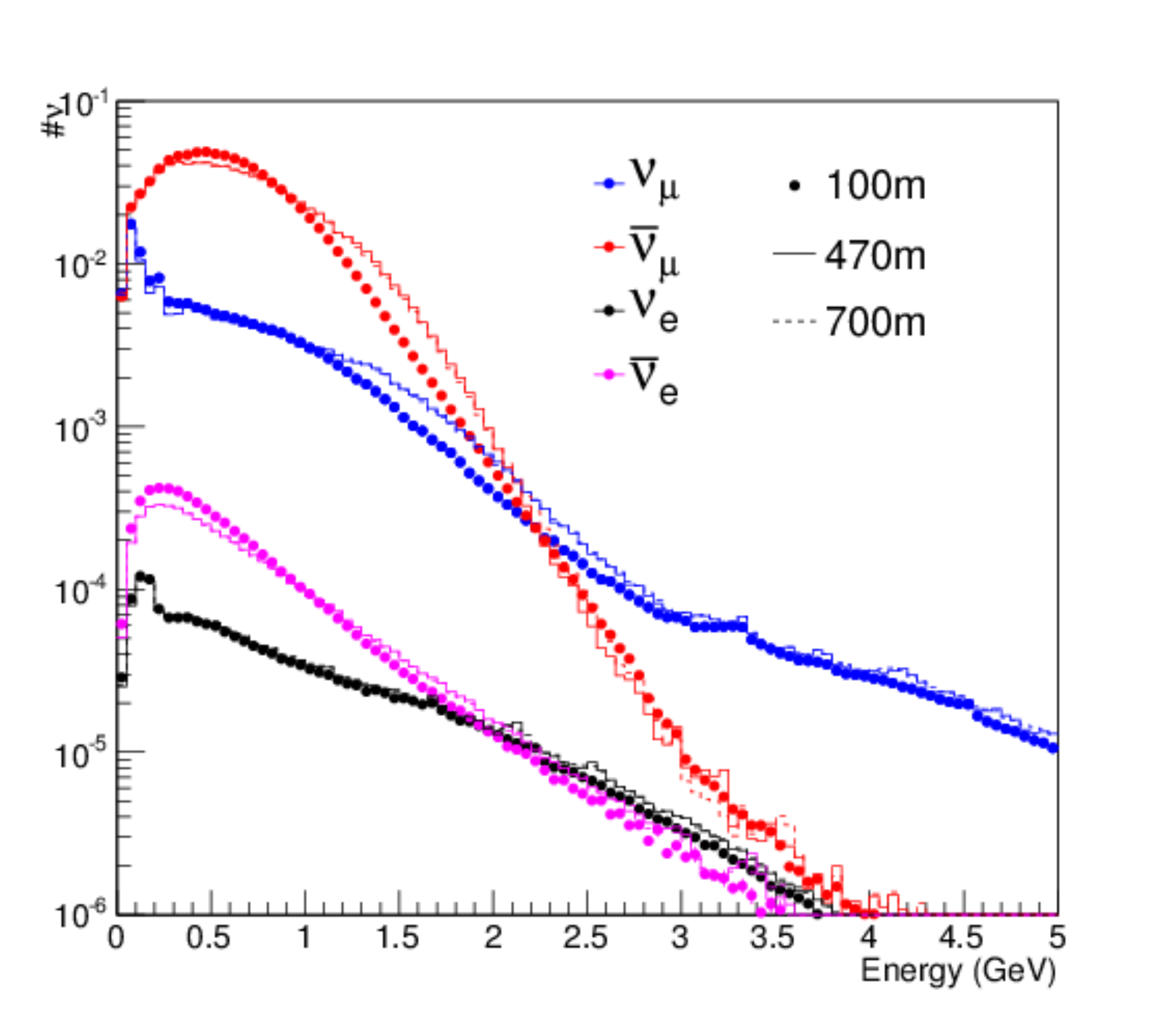}}
\caption{\small Comparison of neutrino fluxes at 100~m, 470~m and 700~m in neutrino (left) and 
antineutrino (right) mode. The total neutrino flux at each location has been normalized to 1 in order to compare the flux shapes. }
\label{fig:fluxes_L}
\end{figure}

\begin{figure}[t]
\centering
\mbox{ \includegraphics[width=0.5\textwidth]{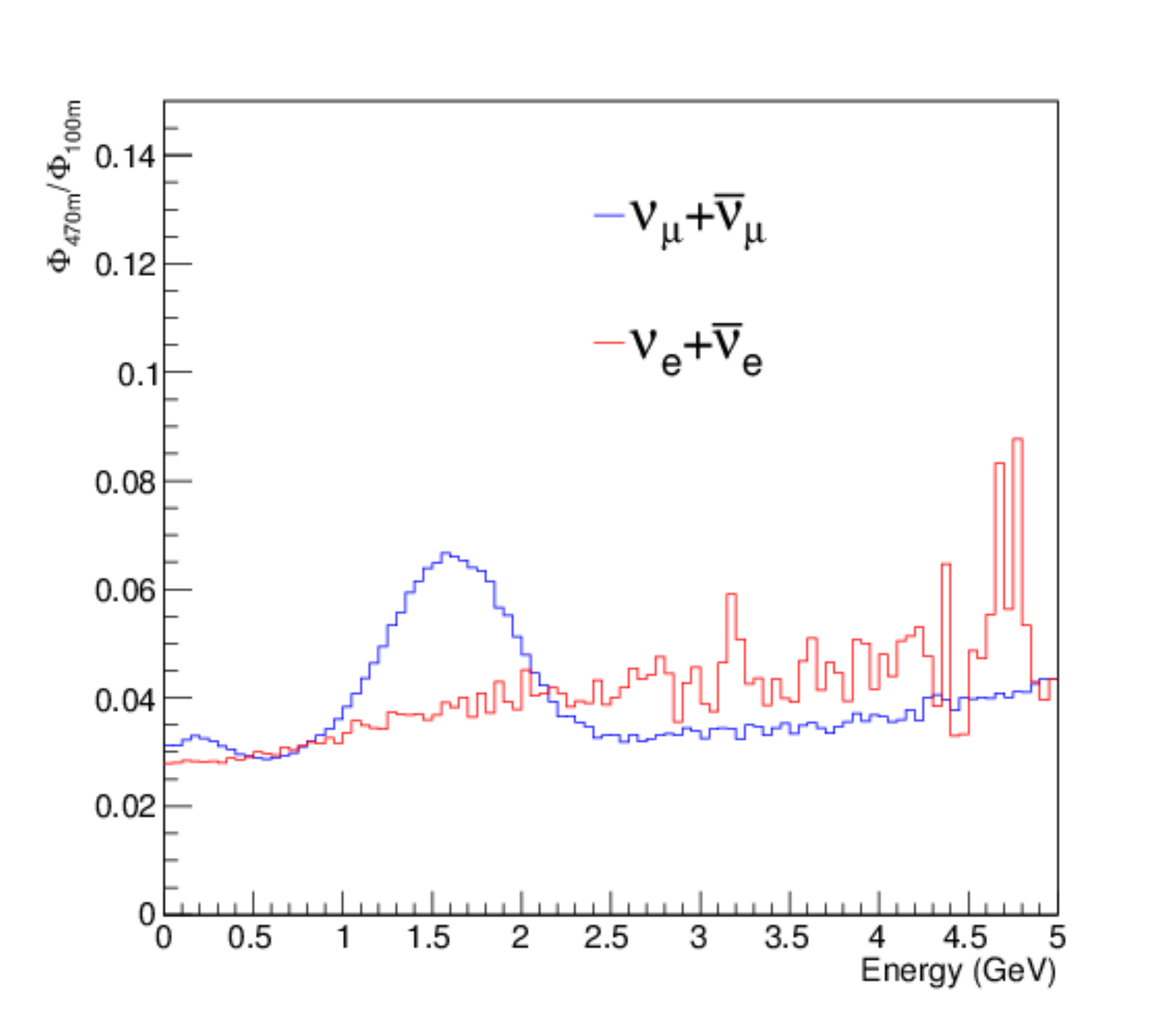} \quad
\includegraphics[width=0.5\textwidth]{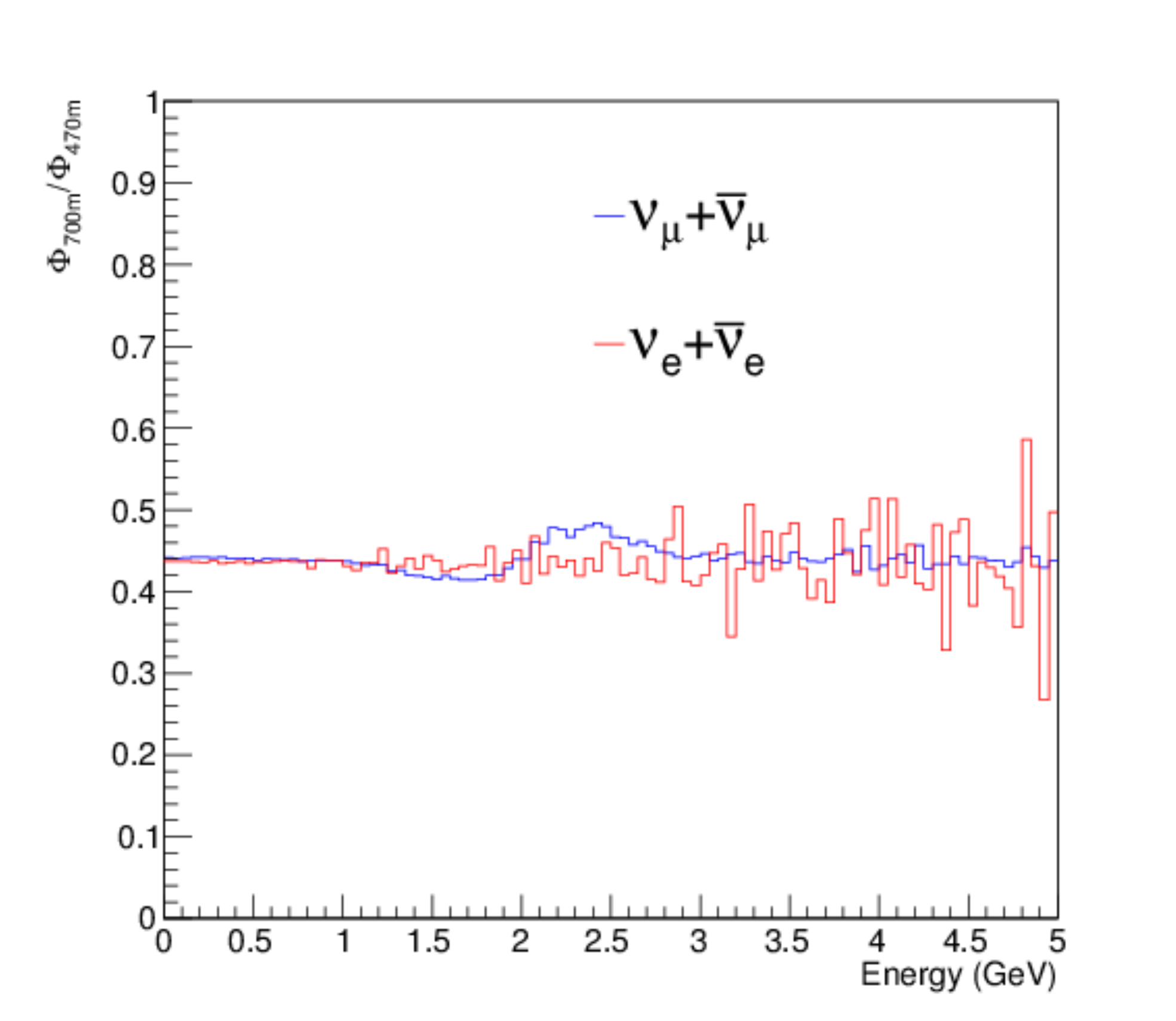}}
\caption{\small Ratios of the muon type and electron type neutrino fluxes at different detector locations.  The left panel compares the flux at \uboone compared to \larnd at 100~m.  The right panel compares the flux at \uboone to another on-axis location at 700~m.}
\label{fig:fluxes_ratio}
\end{figure}

On-axis neutrino fluxes at the BNB are well known; calculations are made with a Geant4-based Monte Carlo simulation \cite{bnb_flux} provided by the \MB collaboration that has been tuned using dedicated hadron production data from the \harp experiment \cite{harp} as well as the world data on kaon production.    Figure \ref{fig:fluxes_L} compares the neutrino fluxes at each of the proposed detector locations in the neutrino and antineutrino beam configurations of the BNB and Fig. \ref{fig:fluxes_ratio} provides the ratios of the absolute fluxes as a function of neutrino energy in order to highlight differences in the shape of the neutrino spectra.  The ratio $\Phi(\uboone)/\Phi(\larnd)$ is on the left and shows clearly the effect of \larnd being only 50~m from the end of the decay region.  The effect is more pronounced in the \numu spectrum, but note that the \nue flux does fall off slightly faster than $1/r^2$ between these two locations $(100^2/470^2 = 0.045)$.   In contrast, the ratio $\Phi(\larfd)/\Phi(\uboone)$ on the right indicates that the neutrino flux past \uboone falls off almost exactly as $1/r^2$ at all neutrino energies $(470^2/700^2 = 0.45)$.  

\subsection{Detector Simulations}
\label{sec:sim}

To estimate physics sensitivities of the experiment, a full Monte Carlo simulation is used.  Neutrino interactions on argon are simulated using the GENIE neutrino event generator \cite{genie} (v2.8.0).  GENIE includes models for all relevant neutrino-nucleon cross sections.  Hadronic interactions of particles as they exit the argon nucleus are simulated using the INTRANUKE package.  Particles exiting the target nucleus are handed off to a Geant4 \cite{geant} simulation within the LArSoft software framework \cite{larsoft} which handles all electromagnetic and hadronic interactions within the liquid argon detector volumes.  Geometry descriptions of the proper dimensions are used for each detector studied.      





\subsection{Sensitivity to \MB Low-Energy \nue Excess}
\label{sec:lee}

In this section we set out to estimate the ability of \larnd to test the nature of the \MB neutrino anomaly in a way independent of any specific oscillation model.  With this analysis, we are addressing the straightforward question: \emph{Does the anomalous excess of electromagnetic events reported by \MB, whether electrons or photons, appear over a distance or exist intrinsically in the beam?}   To answer this we need to make a measurement at a near location in the beam that has good sensitivity to a \MB-like excess.  We set out, therefore, to determine the significance with which \larnd would observe an event excess of the same size as that reported by \MB.  

Starting from the event distribution in the left panel of Fig. \ref{fig:mboone} and reported in \cite{mbosc2}, we directly scale (up or down) the exact event rates observed by \MB, accounting for anticipated differences in event selection efficiencies due to the \lartpc technology as well as fiducial masses, beam exposures and detector locations.  The approach taken in this study does not use a full Monte Carlo simulation (but does rely on MC studies for efficiencies), but the procedure is easy to follow and it is informative to see directly how the technology impacts aspects of the measurement.    
In later sections we will use a full simulation to assess sensitivities to \nue appearance and \numu disappearance with \larnd. 
 
Tables \ref{tab:events} and \ref{tab:errors} illustrate this exercise for the \larnd and \uboone detector configuration.    
The \nue charged-current quasi-elastic (CCQE) selection efficiency in \MB was $\sim$30\%; we conservatively assume a factor 2 increase with \lartpc detectors.  Neutral-current events with single photons in the final state present an irreducible background in \MB since photons look identical to electrons in a Cherenkov detector, but in a \lartpc we estimate a 94\% rejection rate using the $dE/dx$ tag in the first few centimeters of the electromagnetic shower.  

We also scale for differences in detector mass and beam exposure.  For \uboone we assume a 61.4 ton fiducial mass and the approved $6.6 \times 10^{20}$ protons on target (POT).  These are to be compared to the 450 ton fiducial mass of the \MB detector and the $6.46 \times 10^{20}$ POT exposure of the final \MB neutrino mode sample.  Also, the \MB detector was located 541~m from the BNB target while the \uboone detector is a little closer at 470~m.  Simulation confirms that the flux falls off as $1/r^2$ between these two positions.  This gives us the following relative scale factors for electron and single photon final states in \uboone relative to \MB: 
\begin{eqnarray}
f_{\mu B}^{e} & = & 2\times(61.4/450)\times(541^2/470^2)\times(6.6/6.46) = 0.369   \\ 
f_{\mu B}^{\gamma} & = & (0.06/0.30)\times(61.4/450)\times(541^2/470^2)\times(6.6/6.46) = 0.037 \nonumber  
\end{eqnarray}

For \larnd, these scale factors are reduced for the smaller fiducial mass and shorter exposure.  We assume one year of concurrent running with \uboone, or $2.2 \times 10^{20}$ POT.  The neutrino flux, however, is higher at the 100~m location compared to the \uboone location at 470~m.  Figure \ref{fig:fluxes_ratio} shows the \larnd/\uboone flux ratio as a function of neutrino energy according to the beam simulation.  For neutrino energies below $\sim$1~\GeV, there is about 30$\times$ more flux in the near detector.  We use this factor 30 to estimate the number of events in \larnd by scaling from our estimates for \uboone:   
\begin{eqnarray}
f_{ND}^{e} & = & f_{\mu B}^{e} \times (40/61.4)\times(2.2/6.6) \times30 = 2.41 \\ 
f_{ND}^{\gamma} & = & f_{\mu B}^{\gamma}\times (40/61.4)\times(2.2/6.6) \times 30 = 0.24 \nonumber 
\end{eqnarray}

%
In Table \ref{tab:events} we estimate the number of events in nine different background categories by scaling from the \MB neutrino mode results in the energy region below 475~\MeV.   Backgrounds include both intrinsic sources of \nue as well as single photon final state \numu interactions.  We also scale the excess event counts reported by \MB \emph{as if they are electrons} and we refer to this as the "Excess" or the "Signal".  We estimate 49 background events and 47 signal events in \uboone.  In \larnd, with 40 ton fiducial mass and $2.2 \times 10^{20}$ POT exposure, we expect 310 signal events on top of 322 intrinsic \nue and single photon background events.    
 

\begin{table}[t]
   \begin{center}
   \begin{tabular}{| c |  c | c | c || c | c || c | c |}
  \hline
   Process & 200 - 300 & 300 - 475 & Total & Scaling & Total & Scaling & Total \\
   	& MeV (mB) & MeV (mB) &  (mB) & ($\mu$B) & ($\mu$B) &  (LAr1-ND) &  (LAr1-ND)\\
   \hline
   \hline
   \multicolumn{8}{|c|}{Background from intrinsic $\nu_e$} \\
   \hline
   $\mu \rightarrow \nue$		& 13.6	& 44.5	& 58.1	& .369	& 21.5	& 2.40	& 139.8	\\
   $\kplus \rightarrow \nue$	& 3.6		& 13.8	& 17.4	& .369	& 6.4 	& 2.40	& 41.9	\\
   $\kzero \rightarrow \nue$	& 1.6		& 3.4		& 5.0		& .369	& 1.8		& 2.40	& 12.0	\\
   \hline
   \multicolumn{8}{|c|}{Background from $\nu_{\mu}$ misidentification} \\
   \hline
   \numu CC 	& 9.0   	& 17.4	& 26.4	& .185	& 4.9 	& 1.20	& 31.8	\\
   $\numu e \rightarrow
   \numu e$	& 6.1		& 4.3		& 10.4	& .369	& 3.8		& 2.40	& 25.0	\\
   NC $\pi^0$	& 103.5	& 77.8	& 181.3	& .037	& 6.7		& .241 	& 43.6 	\\
   Dirt 		& 11.5	& 12.3	& 23.5	& .037	& 0.9 	& .241	& 5.7		\\
   $\Delta \rightarrow 
   N\gamma $	& 19.5	& 47.5	& 67.0	& .037	& 2.5		& .241	& 16.1		\\
   Other		& 18.4	& 7.3		& 25.7	& .037	& 0.9		& .241	& 6.2		\\   
   \hline
   Background	& 187	& 228	& 415	& 		& {\bf 49.4}	&  		& {\bf 322.1}	\\
   \hline
   Excess 		& 45.2	& 83.7	& 128.9 	& .369 	& {\bf 47.6}	& 2.40	& {\bf 310.2}	\\
   \hline
   \end{tabular}
   \caption{\small Estimated event rates in \uboone and \larnd with reconstructed neutrino energy 200-475~MeV determined by scaling from \MB event rates \cite{mbosc2} and accounting for differences in fiducial masses, beam exposures and selection efficiencies between the detector technologies (see text).}
   \label{tab:events}
   \end{center}
\end{table}

To estimate the significance of a \MB-like signal in \larnd, we apply the fractional systematic uncertainties reported by \MB.  We include an additional 10\% uncertainty on the efficiency of the $dE/dx$ cut applied to separate $e/\gamma$ final states in a \lartpc.  The results of this analysis are shown in Table \ref{tab:errors}.   We find that the 310 signal events expected correspond to a 6.7$\sigma$ excess over the expected background of $322 \pm 18.0\mathrm{(stat)} \pm 42.6\mathrm{(syst)}$, demonstrating that \larnd can verify a \MB-like excess at 100~m with high significance in a very short time.  

\begin{table}[ht]
\begin{center}
   \begin{tabular}{| c |  c | c || c  c |  c || c | c | }
  \hline
   Process & Events 		& Events 		&  \MB		& dE/dx 	& Total	& Error 		& Error \\
   		&  ($\mu$B) 	&  (LAr1-ND) 	&  unc. 	& unc. 	& unc.  	& ($\mu$B) 	& (LAr1-ND) \\
   \hline
   \hline
   $\mu \rightarrow \nue$		& 21.5	& 139.8	& 0.26 	& 0.1		& 0.28	& 6.0		& 39.0		\\
   $\kplus \rightarrow \nue $	& 6.4		& 41.9	& 0.22 	& 0.1		& 0.24 	& 1.55	& 10.1		\\
   $\kzero \rightarrow \nue $ 	& 1.8		& 12.0	& 0.38	& 0.1		& 0.39	& 0.73	& 4.73		\\
   \numu CC 	& 4.9		& 31.8	& 0.26	& 0.0		& 0.26	& 1.27	& 8.26		\\
   $\numu e \rightarrow
   \numu e$	& 3.8		& 25.0	& 0.25	& 0.1		& 0.27	& 1.03	& 6.74		\\
   NC $\pi^0$	& 6.7		& 43.6	& 0.13	& 0.1		& 0.16	& 1.10	& 7.16		\\
   Dirt 		& 0.9		& 5.7		& 0.16	& 0.1		& 0.19 	& 0.16	& 1.07		\\ 
   $\Delta \rightarrow 
   N\gamma $	& 2.5		& 16.1	& 0.14	& 0.1		& 0.17	& 0.43	& 2.77		\\
   Other		& 0.9		& 6.2		& 0.25	& 0.1		& 0.27	& 0.26	& 1.67		\\   
   \hline
   Total		& 49.4	& 322.1	& 	\multicolumn{3}{c||}{}		& 6.55	& 42.6		\\
   \hline
   \hline
   \multicolumn{4}{| r |}{} & \multicolumn{2}{c |}{\uboone} & \multicolumn{2}{c |}{LAr1-ND} \\ 
   \hline
   \multicolumn{4}{| r |}{Total Events} & \multicolumn{2}{c |}{97} & \multicolumn{2}{c |}{632} \\ 
   \multicolumn{4}{| r |}{''Low-energy Excess''} & \multicolumn{2}{c |}{47.6} & \multicolumn{2}{c |}{310.2} \\ 
   \multicolumn{4}{| r |}{Background} & \multicolumn{2}{c |}{49.4} & \multicolumn{2}{c |}{322.1} \\ 
   \multicolumn{4}{| r |}{Statistical Error} & \multicolumn{2}{c |}{7.0} & \multicolumn{2}{c |}{18.0} \\ 
   \multicolumn{4}{| r |}{Systematic Error} & \multicolumn{2}{c |}{6.6} & \multicolumn{2}{c |}{42.6} \\ 
   \multicolumn{4}{| r |}{Total Error} & \multicolumn{2}{c |}{9.6} & \multicolumn{2}{c |}{46.3} \\ 
   \hline
   \multicolumn{4}{| r |}{Statistical Significance of Excess} & \multicolumn{2}{c |}{6.8 $\sigma$} & \multicolumn{2}{c |}{17.3 $\sigma$} \\ 
   \multicolumn{4}{| r |}{Total Significance of Excess} & \multicolumn{2}{c |}{\bf{5.0 $\sigma$}} & \multicolumn{2}{c |}{\bf{6.7 $\sigma$}} \\ 
   \hline
   \end{tabular}
   \caption{\small Anticipated event rate and errors in \uboone and \larnd with reconstructed neutrino energy 200-475~MeV determined by scaling from the observed events at \MB \cite{mbosc2}.  Estimated significance to the \MB ``low-energy anomaly'' at each location is determined by assuming the \MB excess is not dependent on the neutrino propagation length.  
   }
   \label{tab:errors}
   \end{center}
\end{table}

Several things are worth noting about this analysis.  First, the \MB analysis selected charged-current quasi-elastic  events in order to be able to make a reasonable estimate of the neutrino energy in the event by assuming quasi-elastic kinematics in the interaction.  This scaling procedure would, therefore, correspond to only utilizing the \nue CCQE interactions in \larnd.  In a tracking calorimeter detector like a \lartpc it is possible to use the inclusive charged-current event sample, thus increasing further the signal to background ratio in the case of an excess of \nue events.   The analysis presented in this section is conservative in that it only scales the quasi-elastic event rate reported by \MB.  Second, because this analysis does not use a full Monte Carlo simulation, it captures the general features of the expected event rates and their uncertainties but certain details such as the difference in detector volume shapes or the differences between neutrino interactions on carbon vs. argon are ignored.  Nonetheless, the exercise is valuable for estimating the experiment's sensitivity to the exact anomalous event excess reported by \MB, an important aspect of the physics program of \larnd.  In the following sections we will turn to utilizing a full simulation to assess the sensitivity to various searches for oscillations with the combination of \larnd and \uboone.         


\subsection{\numunue Appearance}
\label{sec:nueapp}

In this section we present the sensitivity of an experimental search for sterile neutrinos in the context of a 3 active + 1 sterile neutrino model (3+1).  The signal is taken to be the appearance of electron neutrinos through \numunue transitions according to the two neutrino oscillation probability formula

\begin{equation}
P_{\numunue}^{3+1} = \sin^{2}2\theta_{\mu e} \sin^{2} \frac{\Delta m^2_{41} L}{4E}
\label{eq:app}
\end{equation}

\noindent where $\Delta m^2_{41}$ is the mass splitting between the known mass states and a new state with $\Delta m^2_{41} >> \Delta m^2_{31}$.  Sin$^{2}2\theta_{\mu e}$ is an effective mixing amplitude that combines the amount of mixing of \numu and \nue with mass state $\nu_4$

\begin{equation}  
\sin^{2}2\theta_{\mu e} \equiv 4|U_{\mu 4} U_{e4}|^2
\end{equation}

To assess the sensitivity to appearance signals, we calculate a simple \chisq between the expected background events and the total event rate predicted for a combinations of oscillation parameters $\dmsq_{41}$ and $\sinth_{\mu e}$.  
%
%

The observed \nue candidate event rate in \larnd at 100~m is used to predict the expected rate (in the absence of oscillations) in \uboone at 470~m. 
A near detector of the same technology allows cancelations of systematic uncertainties associated with electron identification efficiencies, background mis-identification rates, neutrino fluxes and neutrino-nucleus cross sections.  The lower limit in the systematic uncertainties on the far detector background prediction is determined by the statistical power of the near detector sample, so a large ND sample is important.  To determine the size and spectrum of expected \nue candidate event samples in \larnd and \uboone, we apply the following selections to simulated events in both detectors:   

\begin{itemize}
\item $\mathbf{\nue}$ {\bf CC :} Electron neutrino charged-current interactions occurring within the fiducial volume are accepted with an 80\% identification efficiency. 

\item {\bf NC $\mathbf{\pizero}$ production :} For neutral-current interactions with any number of \pizero in the final state, the decay photons are analyzed.  If more than one photon converts within the fiducial volume, the event is rejected.  For events where only one photon converts within the fiducial volume, a 94\% photon rejection rate is applied. This is the assumed efficiency of the $dE/dx$ cut to separate $e/\gamma$ in the \lartpc detector.   Note that further rejection of this class of events is likely possible by identifying the low-energy hadronic debri near the vertex of the neutral-current interaction.  The electromagnetic shower produced by the photon will be separated from, but point back to, this vertex.  An observed gap between the photon shower and the interaction vertex would allow further rejection of single photon neutral-current events.  

\item {\bf NC $\mathbf{\gamma}$ production :} Neutral-current interactions resulting in photons in the final state (not from \pizero decays) are analyzed to determine if the photon converts within the fiducial volume.  Again, 6\% of these events are accepted into the \nue candidate sample, corresponding to the expected efficiency of applying $dE/dx$ separation of $e/\gamma$ showers in the \lartpc.  Just as in the photon production from $\pi^0$ described above, there is the additional possibility to reject these events through the identification of a separated interaction vertex which is not being utilized here.

\item {\bf $\mathbf{\numu}$ CC :} For \numu charged-current interactions within the fiducial volumes, we assume 0.1\% are mis-identified as electron neutrino interactions.   Such events can enter into the sample only if there is an identified electromagnetic shower and the muon is not identified. The presence of the muon, of course, tags the event as \numu CC.  
\end{itemize} 

By analyzing the conversion points of photons instead of the true neutrino interaction vertex, we accurately account for acceptance effects in the differently shaped detectors.  Because the $e/\gamma$ separation is performed \emph{entirely with the first few centimeters of a shower}, differences in total shower containment do not affect the assumption that the photon identification efficiency is the same in each detector.  

\begin{figure}[t]
\centering
\mbox{ \includegraphics[width=0.5\textwidth, trim=2cm 0cm 0cm 0cm]{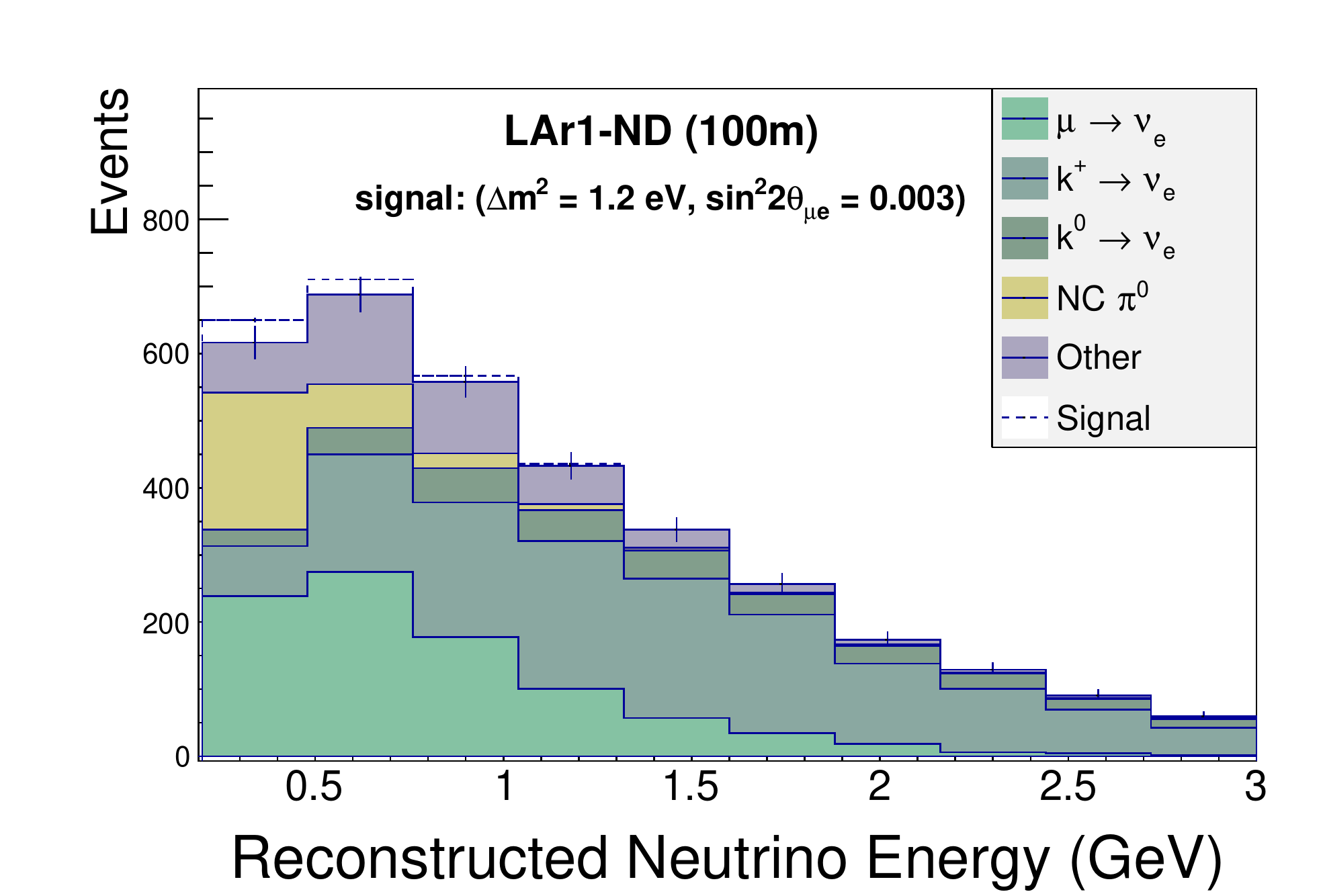} \quad
\includegraphics[width=0.5\textwidth, trim=2cm 0cm 0cm 0cm]{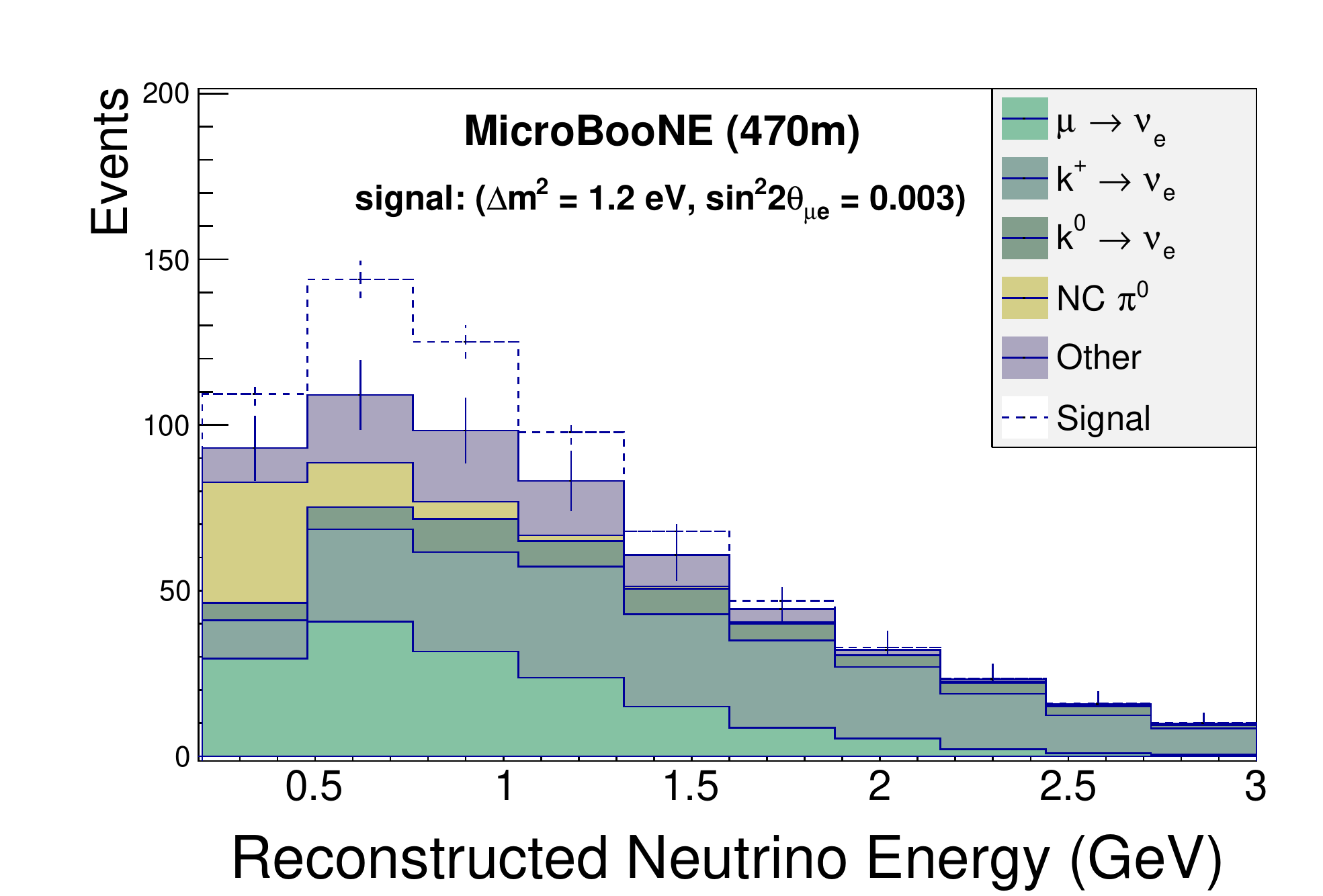}}
\caption{\small Electron neutrino charged-current candidate distributions in \larnd (left) and \uboone (right) shown as a function of reconstructed neutrino energy.  Oscillation signal events for parameter values near the LSND best-fit value are indicated by the dashed blue histograms.}
\label{fig:nue_rates}
\end{figure}

Figure \ref{fig:nue_rates} shows the expected \nue candidate event distributions in \larnd and \uboone for exposures of \ndpot and \ubpot protons-on-target, respectively.   
To simulate a calorimetric energy reconstruction, the neutrino energy in each Monte Carlo event is estimated by summing the energy of the electron candidate and all charged hadronic particles above observation thresholds.  
It should be noted that this method is one possible approach to estimating the neutrino energy in \nue charged-current event candidates. 
The liquid argon TPC technology enables a full calorimetric reconstruction, but other methods can be used as well, such as isolating charged-current quasi-elastic events and assuming QE kinematics.   
The ability to apply complimentary approaches to event identification and energy reconstruction will provide valuable cross checks of the measurements performed.  

\begin{figure}[h!]
\centering
\mbox{ \includegraphics[width=0.5\textwidth, trim=0.5cm 0cm 0cm 0cm]{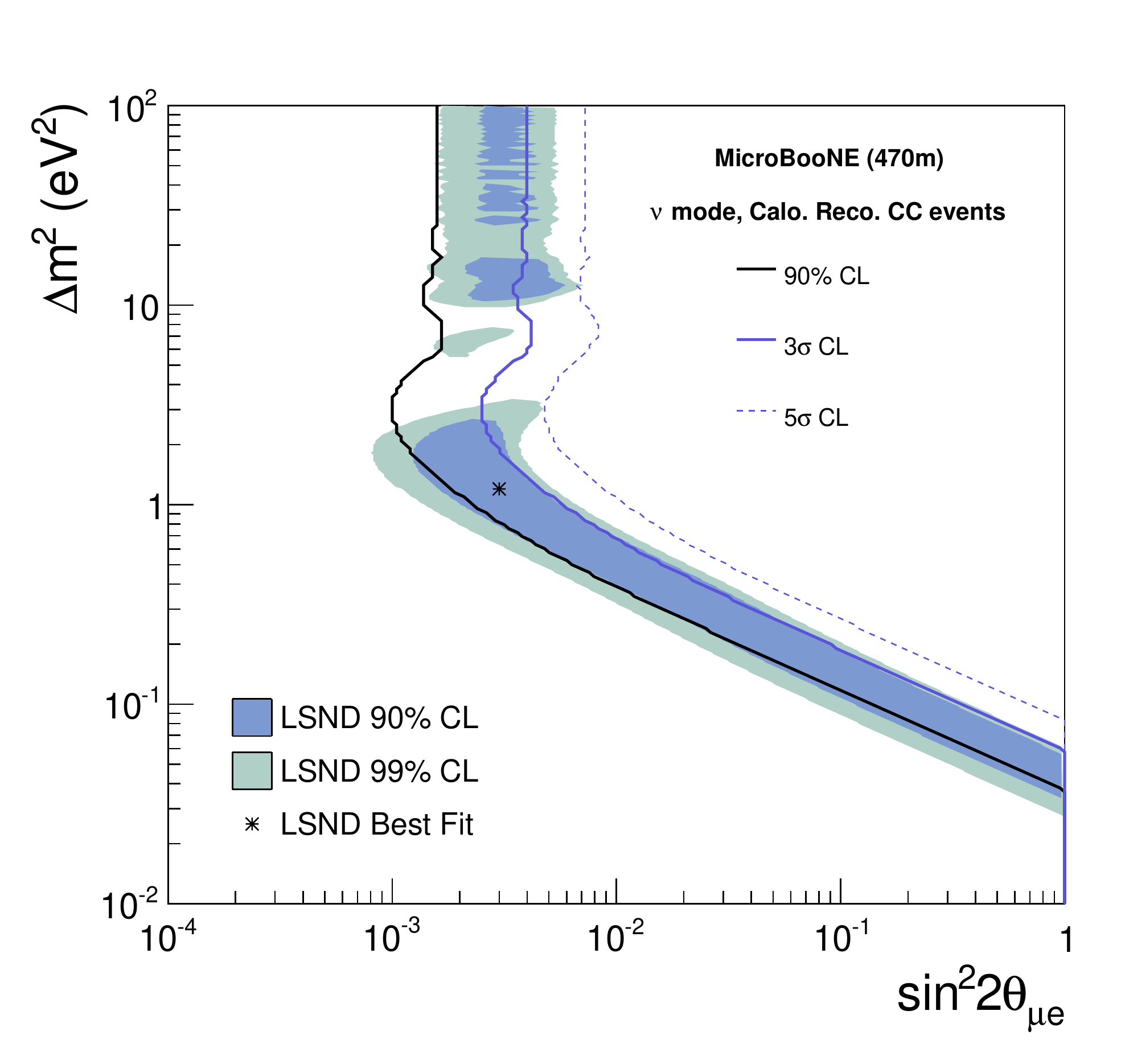} 
 \includegraphics[width=0.5\textwidth, trim=0.5cm 0cm 0cm 0cm]{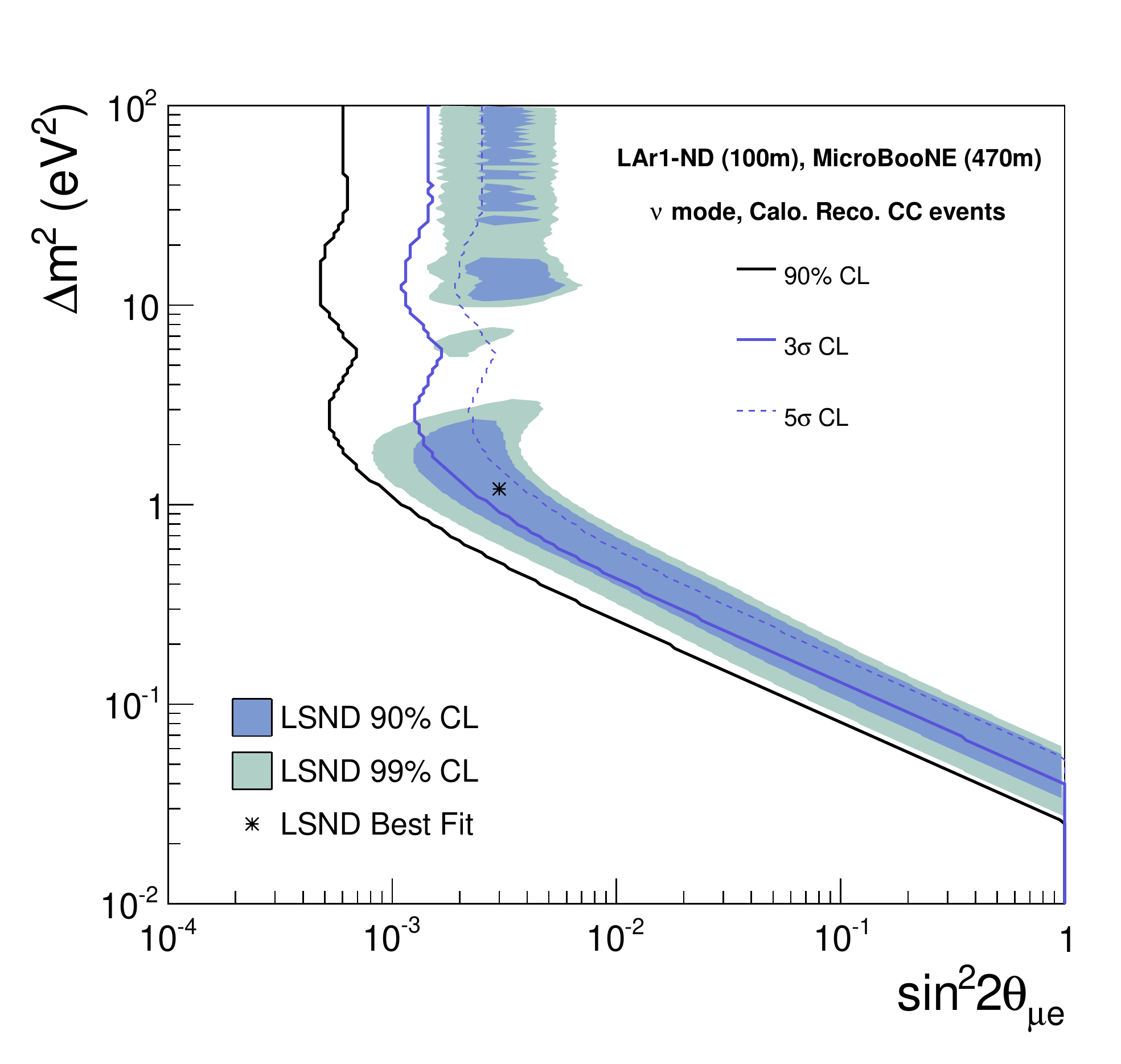} }
\caption{ \small (Left) Sensitivity to \nue appearance in neutrino mode with $6.6 \times 10^{20}$ protons on target exposure for \uboone alone and assuming 20\% systematic uncertainties on \nue backgrounds. (Right) The sensitivity with the same \uboone exposure and including $2.2 \times 10^{20}$ protons on target exposure for \larnd. 
The systematics in the far detector (\uboone) are taken to be the statistical uncertainties in \larnd.}
\label{fig:nueapp_phase1_nobkgd}
\end{figure}

An example signal is included in the event distributions of Fig. \ref{fig:nue_rates}.  The high statistics event sample in \larnd constrains the expected background event rate in \uboone, reducing significantly the systematic uncertainties on the background.       Figure \ref{fig:nueapp_phase1_nobkgd} compares the sensitivity to a \numunue oscillation signal under the 3+1 model of \uboone alone (left) and \uboone + \larnd (right).  The sensitivity is strengthened through the reduction of systematic errors, covering the best-fit point to the LSND data at $\sim4\sigma$.  

\clearpage

\subsection{\numu Disappearance}
\label{sec:numudis}

\begin{figure}[t]
\centering
\mbox{ \includegraphics[width=0.5\textwidth, trim = 1cm 0cm 0cm 0cm]{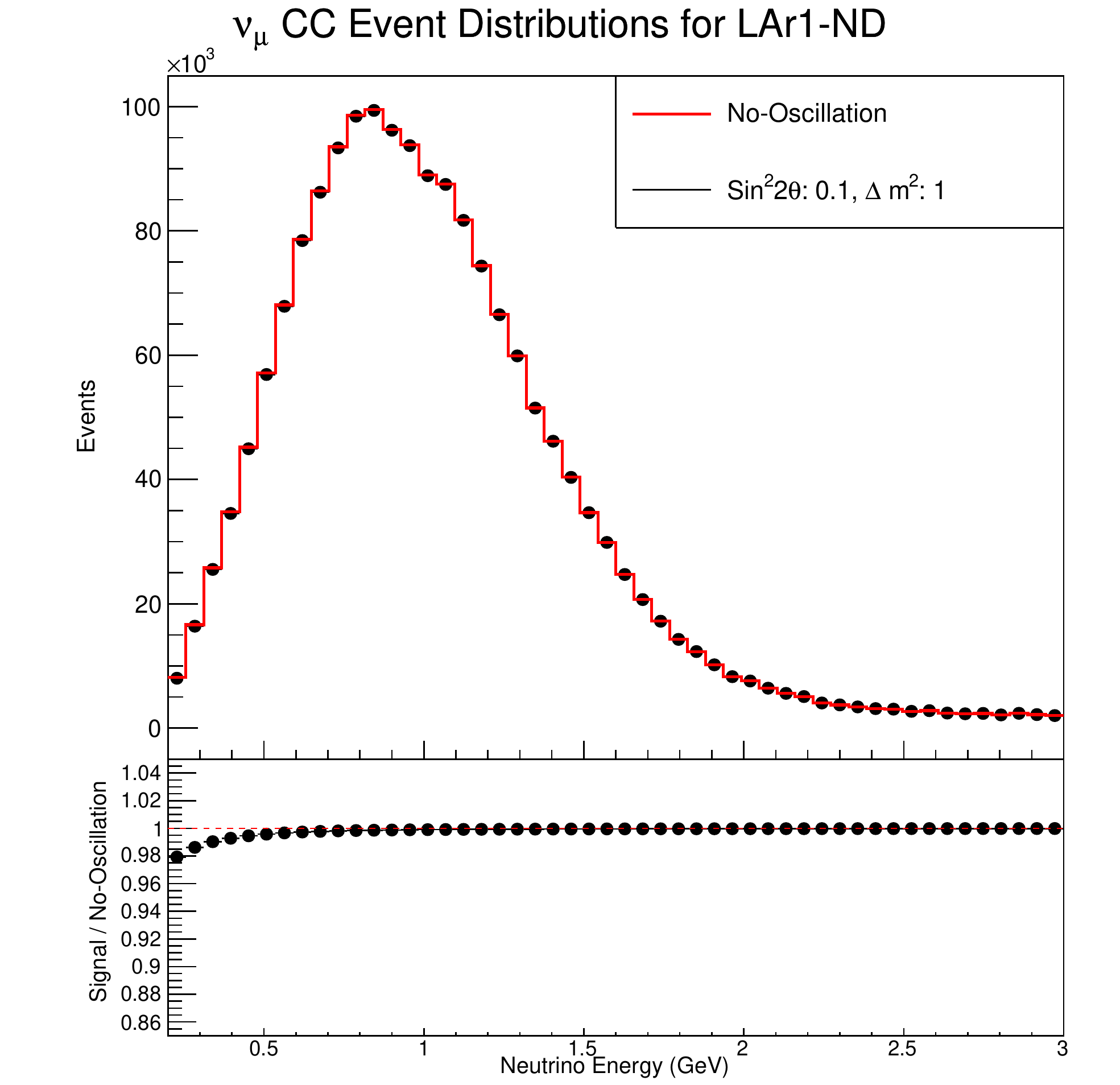} 
\includegraphics[width=0.5\textwidth,  trim = 1cm 0cm 0cm 0cm]{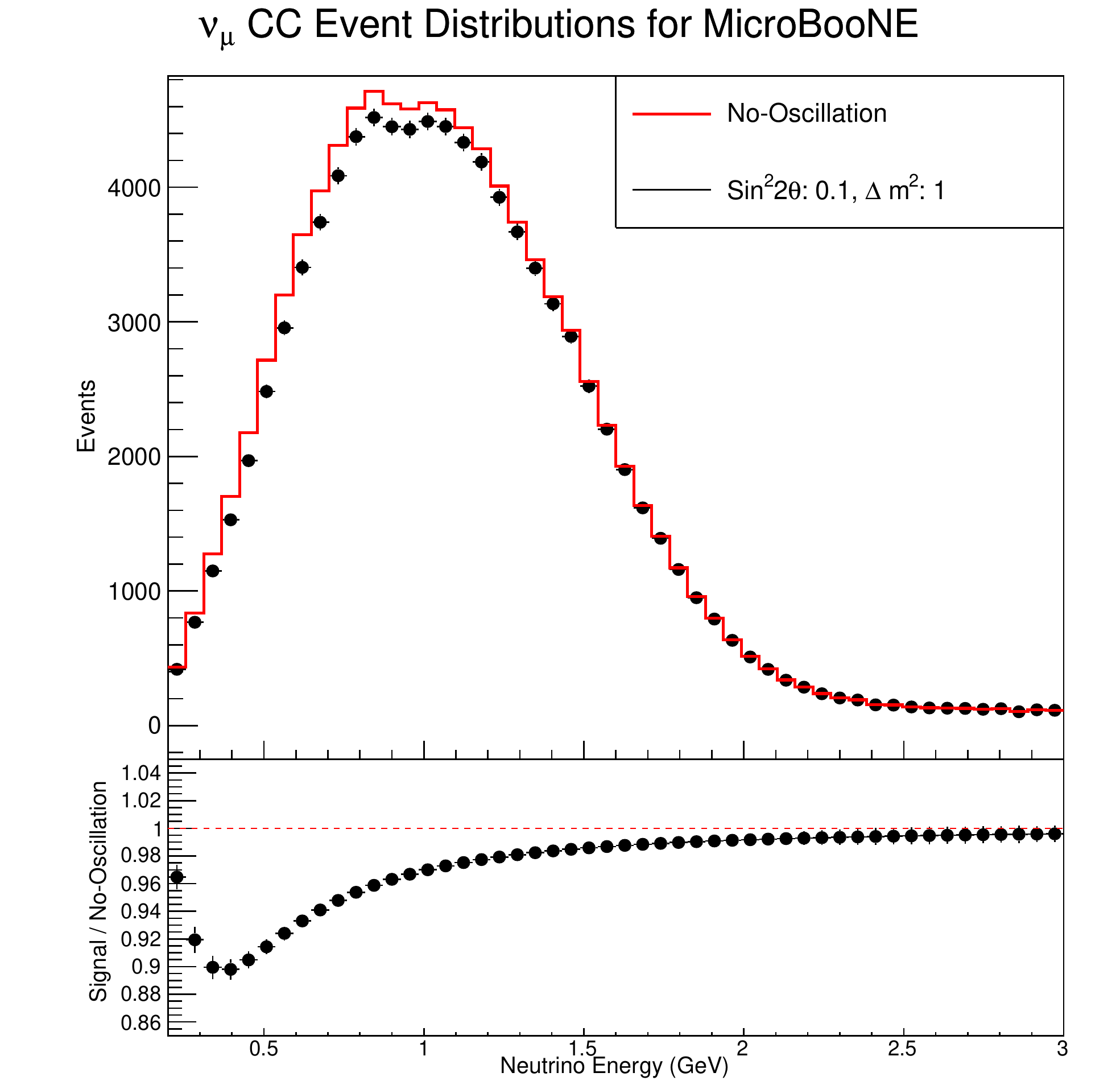} }
\caption{ \small A $\dmsq = 1.0$ \eVsq \numu disappearance signal in \larnd (left) and \uboone (right). Flux and cross section errors of 15-20\% would hide this signal in \uboone alone, but using the observed \larnd spectrum to normalize the expected rate at \uboone makes it observable. }
\label{fig:numudis_events1}
\end{figure}

The disappearance of active flavor interactions over a distance would be a clear signature of oscillations into sterile neutrinos.  Further, an observation of \nue appearance that is to be interpreted as due to an oscillation must be accompanied by a disappearance of muon neutrinos with greater or equal probability.  Therefore, sensitive searches for \numu disappearance are an important component of the test for low-mass sterile neutrinos.  This is a significant element of the physics brought by this proposal because only with a near detector is there the possibility of a sensitive search for \numu disappearance.    

Unlike the search for \nue appearance, which will be limited by small statistics, the search for \numu disappearance in \uboone will be a systematics limited measurement.  Overall normalization errors of 15-20\% impact the \numu charged-current event rate due to uncertainties in both the neutrino flux and interaction cross sections.  \larnd, by providing a high statistics measurement of the \numu interaction rate in the beam (flux $\times$ cross section) before the on-set of any oscillation, enables a search for \numu disappearance at the \uboone location.  
Figure \ref{fig:numudis_events1} shows an example oscillation signal for $\dmsq = 1.0$ \eVsq.   Due to its near location, the effects of the oscillation are barely noticeable in \larnd and only at very low neutrino energies.  At the \uboone location, however, a significant distortion is visible.  The constraint on the predicted rate provided by \larnd enables the observation of this effect.

\begin{figure}[h!]
\centering
\includegraphics[width=0.5\textwidth]{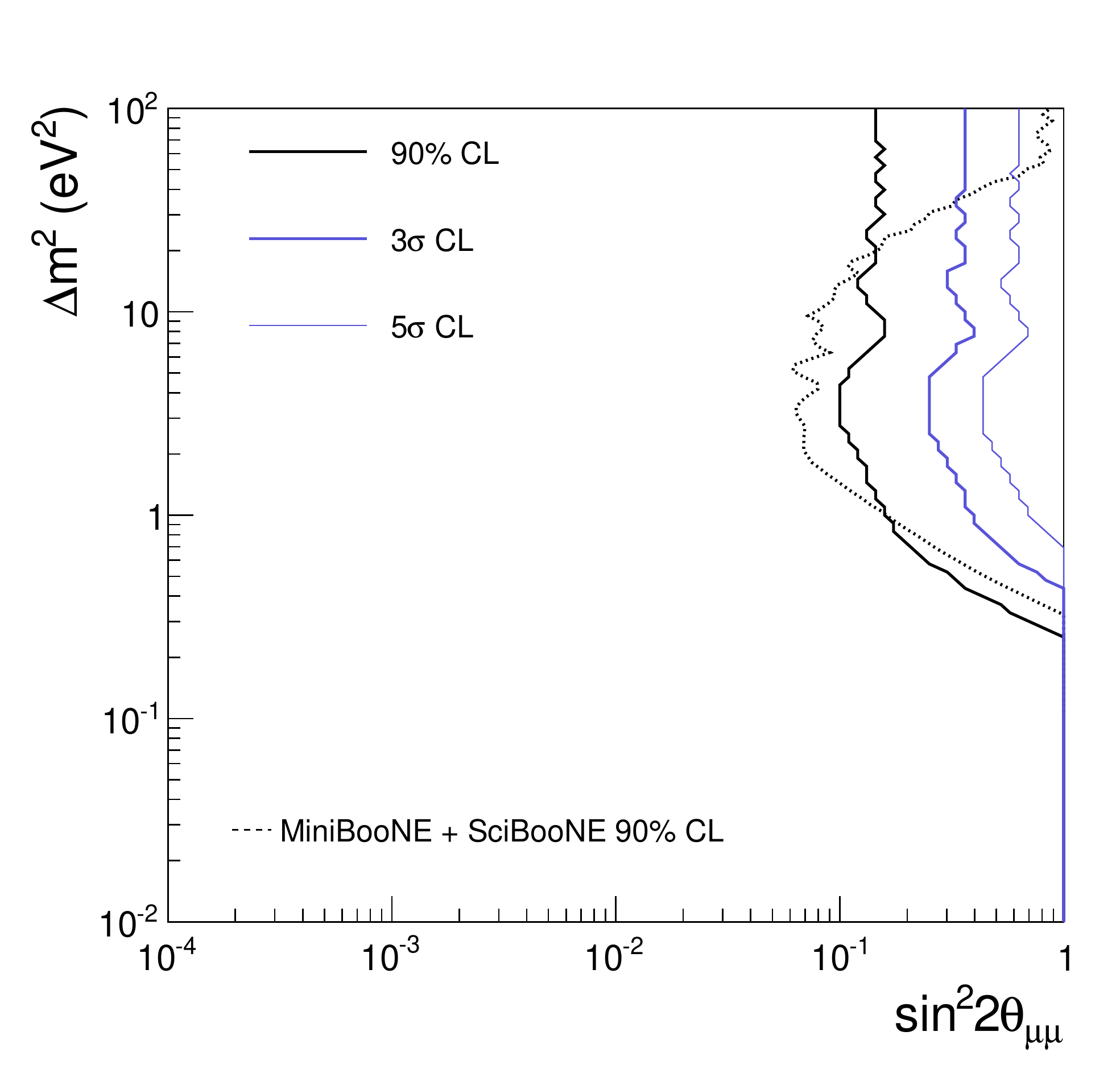} \\
\mbox{ \includegraphics[width=0.5\textwidth]{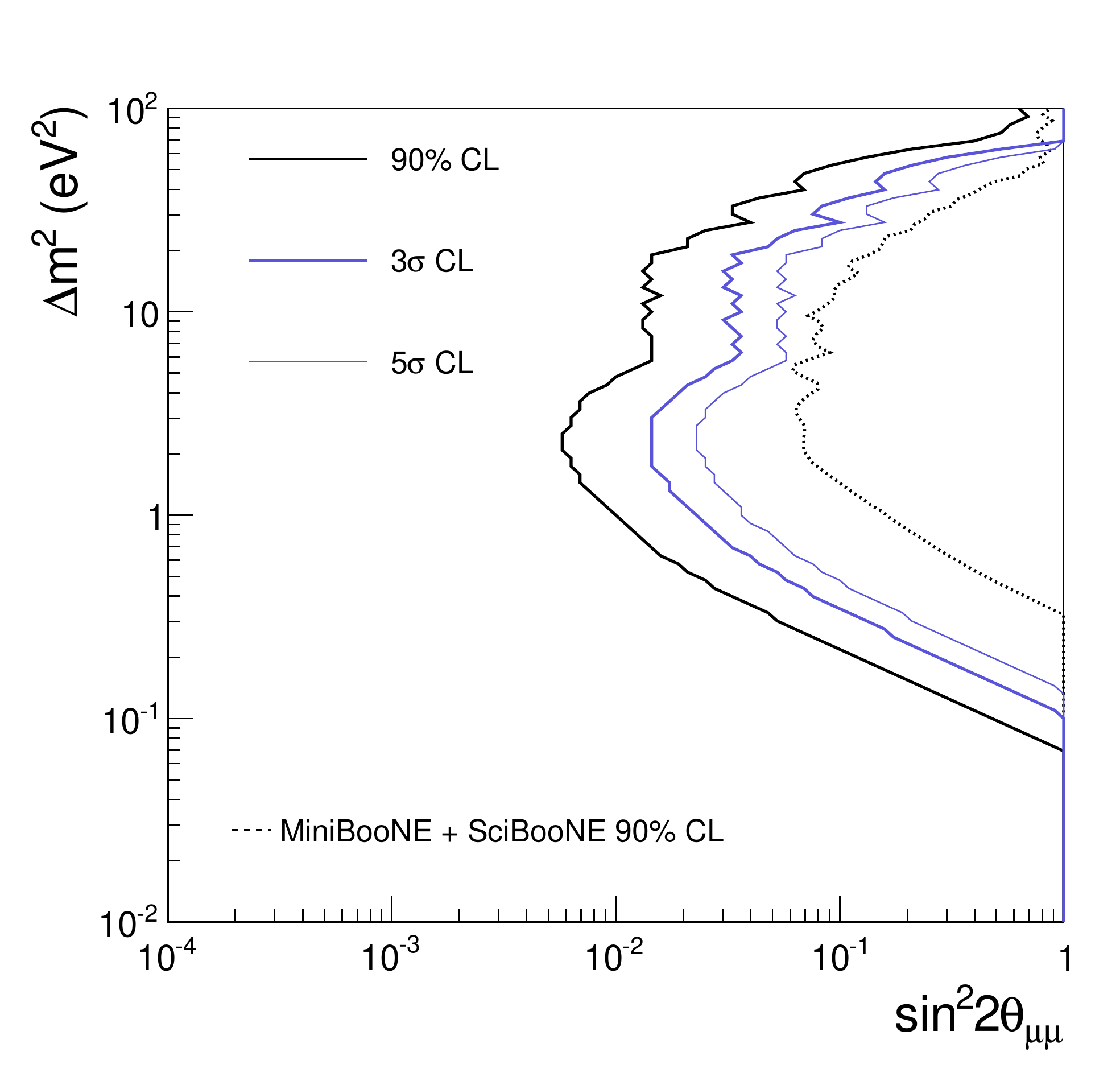}  
\includegraphics[width=0.5\textwidth]{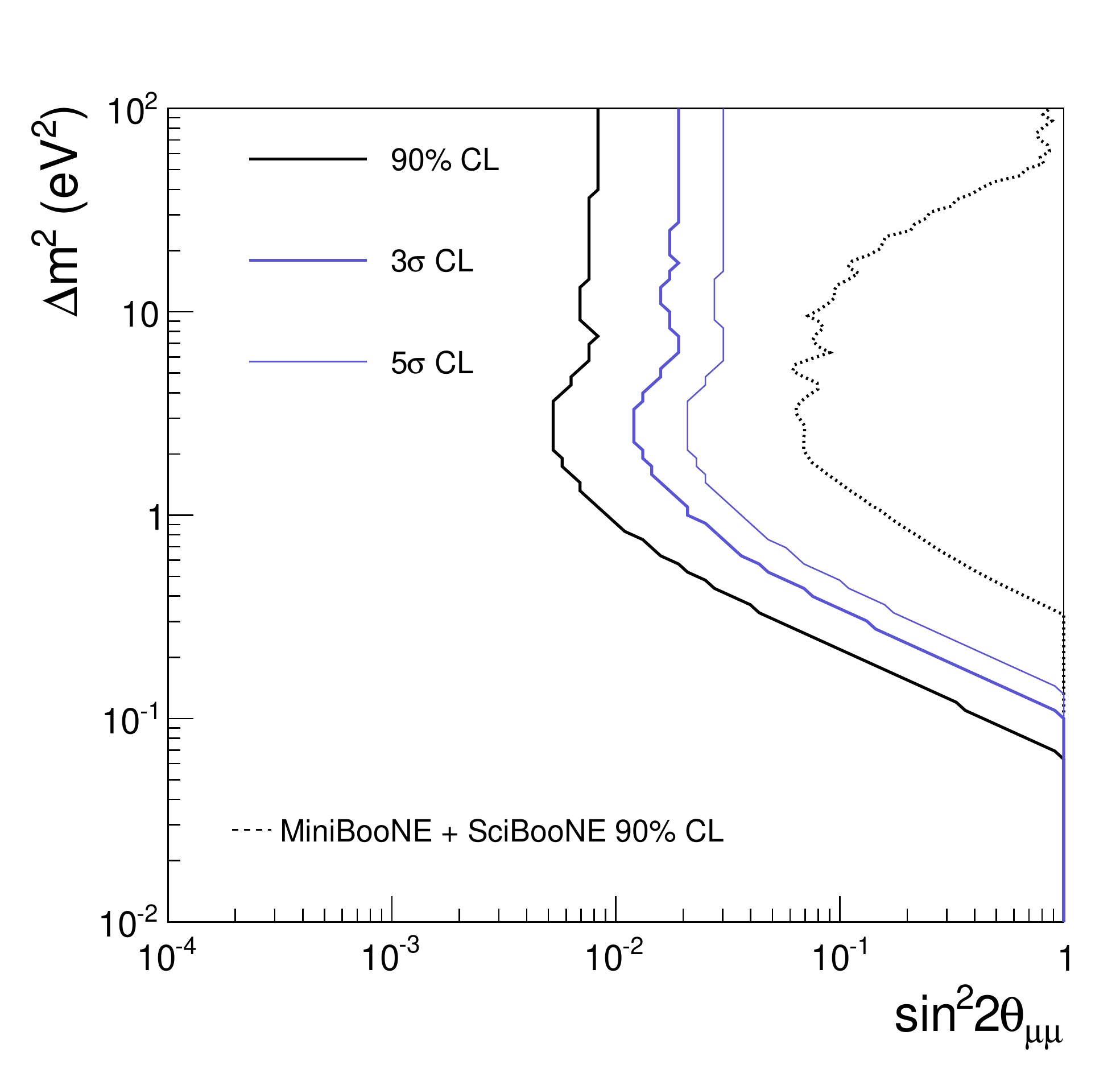} }
\caption{ \small Sensitivity to \numu disappearance of \uboone alone (upper) and including one year of running of \larnd (lower). 
In each case, the dotted line shows the limit set by the SciBooNE + MiniBooNE combined search for \numu disappearance \cite{sciboone} for comparison.  With \uboone alone a 15\% systematic uncertainty on the absolute \numu event rate is assumed.  The lower left panel shows the result of a shape only analysis.  The right panel uses both rate and shape information in the two detectors.}
\label{fig:numudis_phase1}
\end{figure}

Figure \ref{fig:numudis_phase1} demonstrates the effectiveness of this constraint.  The left panel shows the estimated sensitivity to \numu disappearance using \uboone alone.   One can see that \uboone alone is not competitive.  The center and right panels illustrate the improvement with \larnd serving as a near detector. 
The measurement in \larnd is used to normalize the far detector spectrum in the absence of oscillations and the statistical uncertainties in the near detector determine the systematics in the far detector (\uboone).  The center panel shows the result of a \emph{shape only} analysis in the far detector.  The integrated rate observed in \larnd is used to normalize the predicted spectrum at \uboone. 
In this way, any oscillation signal at the near detector is normalized out of the data.  This leads to the reduction in sensitivity at very high \dmsq where rapid oscillations result in a significant reduction in the \numu event rate even at the near detector.  If instead, we assume some sensitivity (15\% normalization uncertainty) also to the reduced rate in the near detector, then a \emph{rate + shape analysis} of the data in both detectors improves the high-\dmsq sensitivity as shown in the right panel of Fig. \ref{fig:numudis_phase1}.  

\clearpage

\subsection{Probing Active to Sterile Oscillations With Neutral-Currents}
\label{sec:ncdis}

A unique probe of sterile neutrino oscillations, directly sensitive to any ``sterile" flavor content, is available through neutral-current (NC) neutrino interactions.  
In this type of search, one looks for an overall depletion of the flavor-summed flux $\times$ cross section event rate. As with \numu disappearance, this type of search is significantly less sensitive in a single-detector experiment due to large systematic uncertainties in flux and cross section. A two-detector experiment provides superior sensitivity by removing these systematic errors.

In an NC-based search, neutrino energy reconstruction is impossible because the surviving neutrino carries off an unknown amount of energy. 
Similarly, Flavor identification is also not possible.
Therefore, one relies on looking for an overall deficit of the flavor-summed event rate. For simplicity, we have considered here the NC $\pi^0$ channel, due to its characteristic event topology and kinematics. Unlike other NC channels, the presence of the two photons from the $\pi^0$ decay pointing back to a common vertex, with an invariant mass corresponding to $m_{\pi0}$, should provide a powerful discriminant against potential backgrounds. This yields a better understood event sample that is less susceptible to systematic errors. 

We expect around 30,000 NC \pizero interactions in the \larnd with 40 tons fiducial volume and \ndpot proton on target exposure.  Assuming a 50\% selection efficiency this corresponds to an approximately 1\% statistical uncertainty on a measurement of the unoscillated NC \pizero event rate.  Similar \lartpc detectors at both locations means that systematic uncertainties will largely cancel in predicting the rate at \uboone, making neutral-current disappearance an excellent and complementary way to test for new physics with \larnd.

\subsection{Anomalous Single Photon Production}
\label{sec:photon}

It is possible the event excess observed by \MB is not due to \numunue oscillations but is instead comprised of an as-yet unknown source of neutral-current interactions producing single photons in the final state.  Cherenkov detectors, such as \MB, are not able to distinguish electromagnetic showers initiated by photons versus by electrons.  \uboone, on the other hand, by applying the \lartpc technology, will determine if the excess at low reconstructed energies in the Booster Neutrino Beam at $\sim$500~m is coming from \nue's or photons produced in (mostly \numu) NC interactions.   In the case of photons, this becomes an important realization since a new source of single photon final states has implications for other Cherenkov detector neutrino experiments in the same energy range, such as T2K.   

In confirming that the observed \MB excess is photons, \uboone would see a few dozen events above expected backgrounds at low-energy.  In this scenario, \larnd at 100~m will allow us to immediately confirm that the excess is intrinsic in the beam (i.e. that it is some standard, but un-modeled neutral-current interaction), and, importantly, \larnd will have hundreds of events per year instead of ten, as shown in Table \ref{tab:events}.   Such a sample will enable a measurement of this unknown reaction with much greater precision and inform the development of cross section models in this energy range to include this process with the correct rate.  This will be important input for some accelerator-based neutrino experiments studying oscillations at the atmospheric \dmsq that observe signals in the few hundred MeV energy range.   

\subsection{Neutrino Cross Sections}
\label{sec:xsec}


Precise cross section measurements are considered a fundamental prerequisite for every neutrino oscillation study.  In the energy range of interest, as a result of competitive physical processes and complicated nuclear effects, neutrino interactions on argon include a variety of final states.  These can range from the emission of multiple nucleons to even more complex topologies with multiple pions, all in addition to the leading lepton.
 
Liquid argon TPC technology is particularly well suited to this purpose because of its excellent particle identification capability and calorimetric energy reconstruction down to very low thresholds.  In addition, \larnd will obtain good muon energy reconstruction from the downstream muon range detector.

\larnd provides an ideal venue to conduct precision cross section measurements in the critical 1~GeV range.  A novel approach based on the event categorization in terms of exclusive topologies can be used to analyze data and provide precise cross section measurements.  Due to its location near the neutrino source (20-30$\times$ the flux at \uboone) and relatively large mass (0.65$\times$ the \uboone mass), \larnd will make measurements of neutrino interactions with high statistics, as shown in Table \ref{tab:events_xsec}.  In the table, we show the expected rate of events in their main experimental topologies.  Included for reference, we also show the classification by physical process from Monte Carlo truth information.  An exposure of one year (\ndpot POT) with \larnd will provide an event sample 5-6$\times$ larger than will be available in \uboone alone.

\begin{table}[h!]
   \begin{center}
   \begin{tabular}{ l l r}
   {\bf Process} &  & {\bf No. Events} \\
   \hline
   	\hline
	 \multicolumn{3}{c}{\emph{\numu ~ Events (By Final State Topology)}} \\
	CC Inclusive 		& & 449,959 \\
   	CC 0 $\pi$~~~~~~~~~~~~ & $\numu N \rightarrow \mu + Np $ 	& 307,441 	\\
			   		& $~\cdot~\numu N \rightarrow \mu  +  0\prot $			& 73,863 	\\
			   		& $~\cdot~\numu N \rightarrow \mu  + 1\prot $ 			&173,830	\\
			   		& $~\cdot~\numu N \rightarrow \mu  + 2\prot $ 			& 29,894 	\\
			   		& $~\cdot~\numu N \rightarrow \mu + \geq3\prot $ 		& 29,854	\\
  	CC 1 $\pi^{\pm}$   		& $\numu N \rightarrow \mu + \rem{nucleons} + 1\pi^{\pm}$ 	& 99,446 	\\
	CC $\geq$2$\pi^{\pm}$ 	& $\numu N \rightarrow \mu + \rem{nucleons}~+ \geq2\pi^{\pm}$ 	& 8,433  	\\
	CC $\geq$1$\pi^{0}$		& $\numu N \rightarrow \rem{nucleons}~+ \geq1\pi^{0}$ 		& 43,048	\\
	\multicolumn{3}{c}{} \\
	NC Inclusive & & 171,869 \\
	NC 0 $\pi$   		& $\numu N \rightarrow \rem{nucleons} $ 		& 118,787  	\\
  	NC 1 $\pi^{\pm}$   		& $\numu N \rightarrow \rem{nucleons} + 1\pi^{\pm}$ 		& 22,407  	\\
	NC $\geq$2$\pi^{\pm}$ 	& $\numu N \rightarrow \rem{nucleons}~+ \geq2\pi^{\pm}$ 	& 2,788 	\\
	NC $\geq$1$\pi^{0}$		& $\numu N \rightarrow \rem{nucleons}~+ \geq1\pi^{0}$ 		& 30,910	\\
   	\hline
	\hline
	\multicolumn{3}{c}{\emph{$\nu_{e}$   Events}} \\
	 CC Inclusive		&  	 & 3,465 \\
	 NC Inclusive		&	 & 1,195 \\
	 \hline
	 \hline
	 Total $\numu$ and $\nu_e$ Events	&		& 626,488 \\

	 	\multicolumn{3}{c}{} \\
	\hline
	\hline
	 \multicolumn{3}{c}{\emph{\numu Events (By Physical Process })} \\
   	CC QE 			& $\numu n \rightarrow \muminus \prot$			& 270,623	 \\
	CC RES			& $\numu N \rightarrow \muminus N $ 		& 124,417	 \\
	CC DIS			& $\numu N \rightarrow \muminus X$ 			& 46,563	\\
	CC Coherent 			& $\numu Ar \rightarrow \mu Ar +  \pi$ 		& 1,664	\\
	\hline
	 \hline

   \end{tabular}
   \caption{\small Estimated event rates using GENIE in a \ndpot POT exposure of \larnd.  In enumerating proton multiplicity, we assume an energy threshold on protons of 21~MeV.  The $0 \pi $ topologies include any number of neutrons in the event.}
   \label{tab:events_xsec}
   \end{center}
\end{table}

\clearpage

\section{LAr1-ND Detector Concept}
\label{sec:DetectorConcept}

\begin{figure}[t]
\centering
\setlength{\fboxsep}{0pt}%
\setlength{\fboxrule}{1.5pt}%
\mbox{ \fbox{\includegraphics[width=0.47\textwidth]{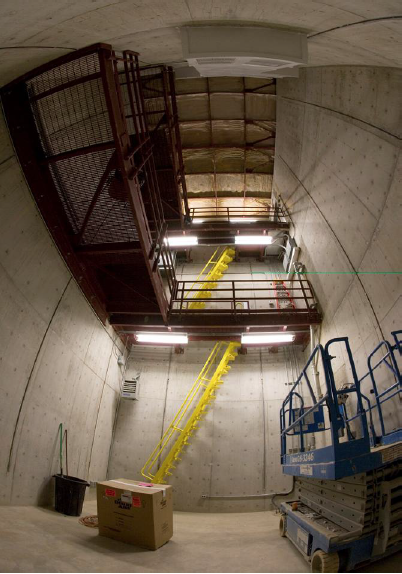}} \quad
\includegraphics[width=0.56\textwidth]{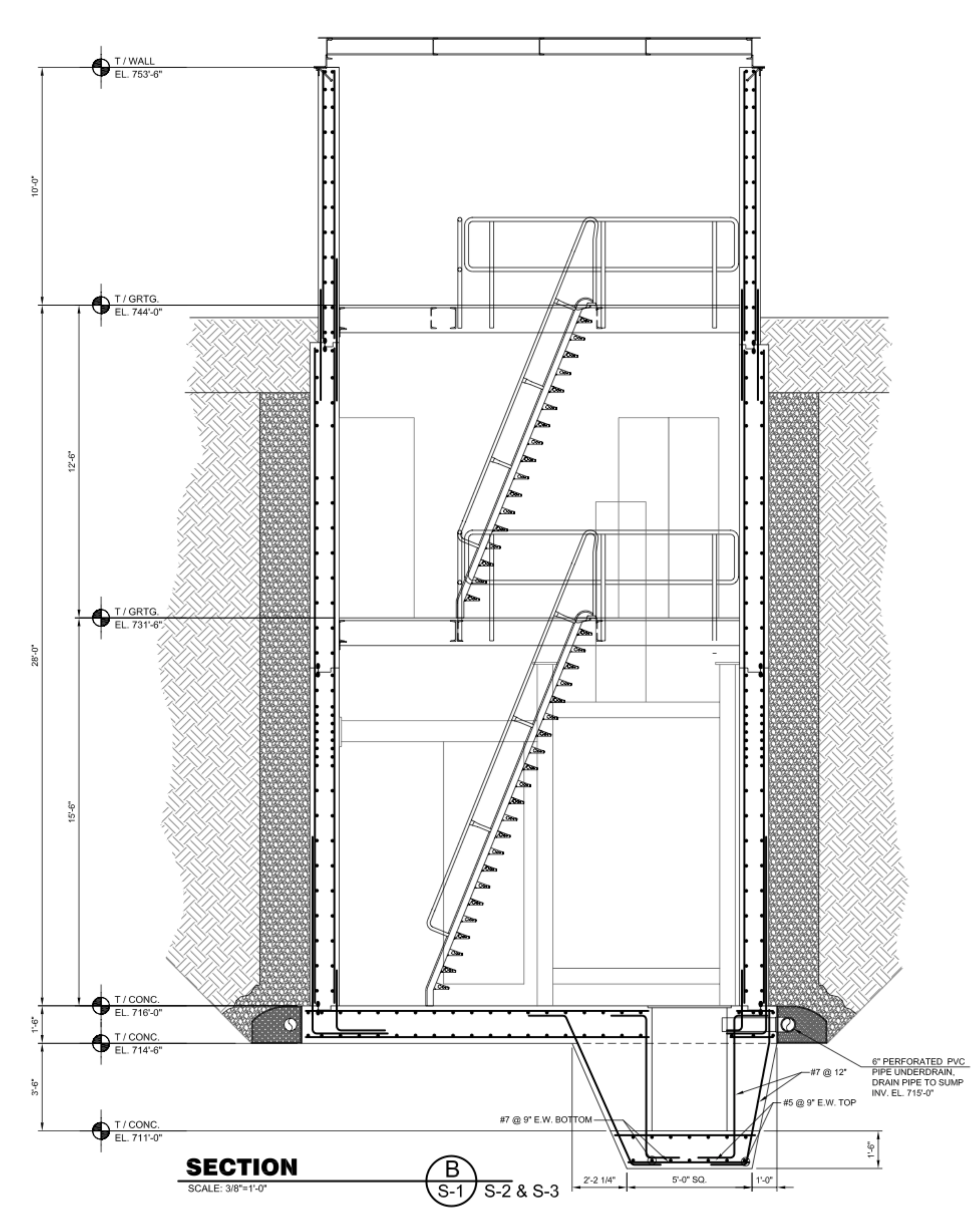}}
\caption{ \small (Left) The empty \SB detector hall located at 100~m along the Booster Neutrino Beam. (Right) Side view of the \SB enclosure.  The neutrino beam enters from the right.}
\label{fig:SciBooNE-pit}
\end{figure}

The \larnd detector, in addition to bringing the physics program outlined above, will serve as a valuable R\&D project for the continued development of the \lartpc technology toward LBNE.  All major systems of the detector can provide important opportunities to test new designs, construction techniques, and physics performance.    
The basic concept for \larnd is a membrane-style cryostat to be constructed in the experimental enclosure previously constructed for the \SB experiment.  Located on-axis at 100 m from the Booster Neutrino Beam target, this existing structure provides a perfect location for a BNB near detector.  The \SB enclosure, shown in Fig. \ref{fig:SciBooNE-pit}, has the following interior dimensions:
\begin{itemize}
\item Length (beam direction) = 4.9~\m
\item Width = 7.0~\m
\item Depth: floor-grade = 8.5~\m, floor-ceiling = 11.6~\m.
\end{itemize}       

\begin{figure}[h!]
\centering
\setlength{\fboxsep}{0pt}%
\setlength{\fboxrule}{1pt}%
\fbox{\includegraphics[width=0.8\textwidth]{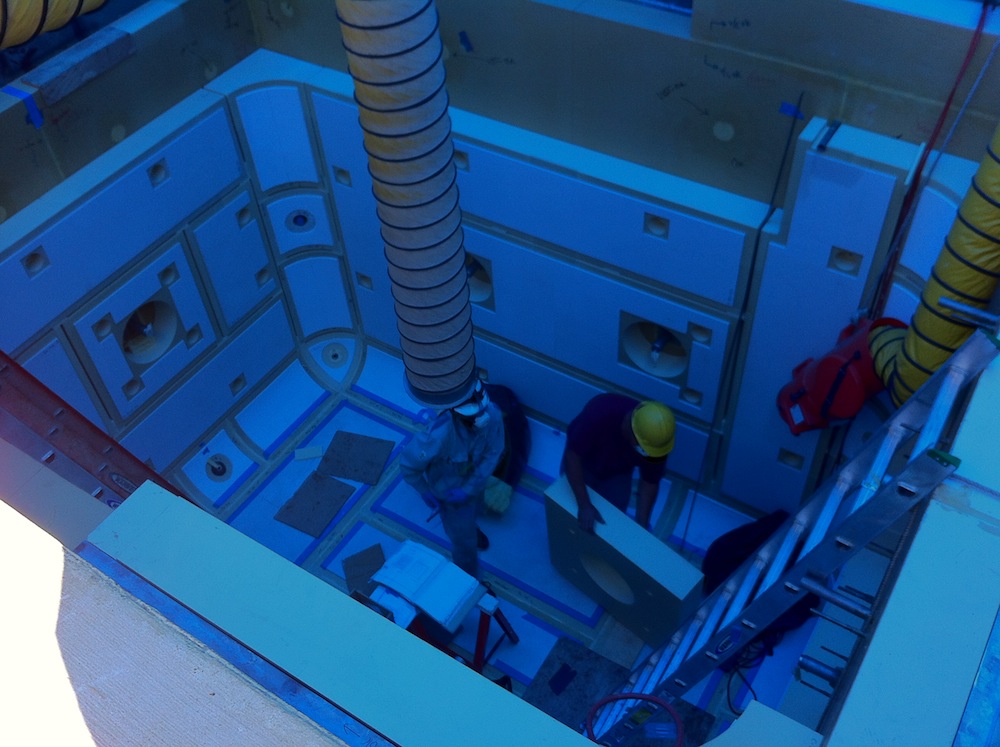}}
\caption{\small The 35 ton prototype membrane cryostat under construction at Fermilab.}
\label{fig:35ton}
\end{figure}

\begin{figure}[h]
\centering
\mbox{ \includegraphics[width=0.9\textwidth]{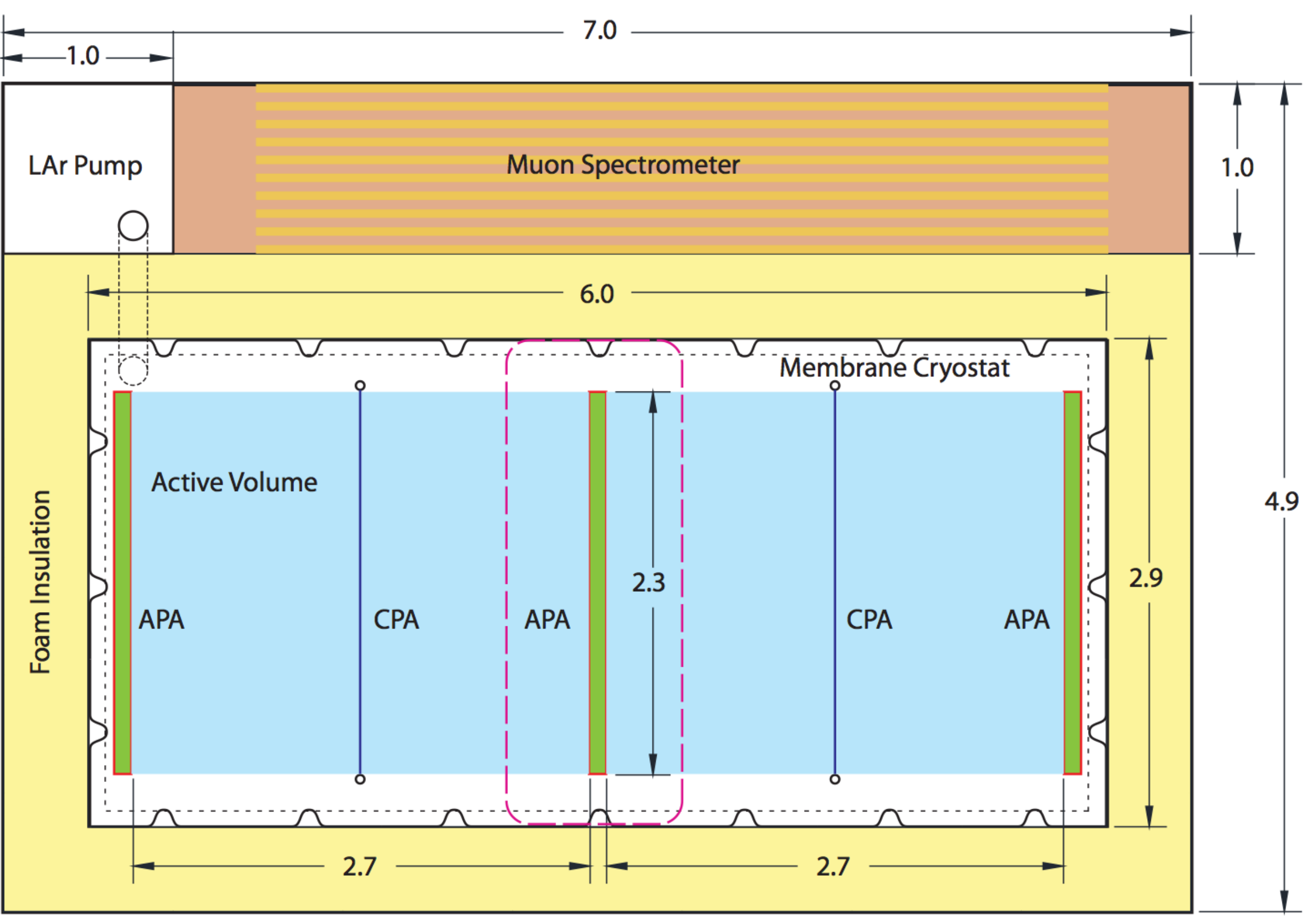} }
\caption{\small Top view schematic drawing of the \larnd detector concept.  A membrane cryostat construction is built to fill the existing enclosure.  0.5~m of foam insulation surrounds the corrugated steel membrane.  The neutrino beam enters from the bottom in this graphic.}
\label{fig:larnd_det1}
\end{figure}

\begin{figure}[h!]
\centering
\includegraphics[height=0.8\textheight]{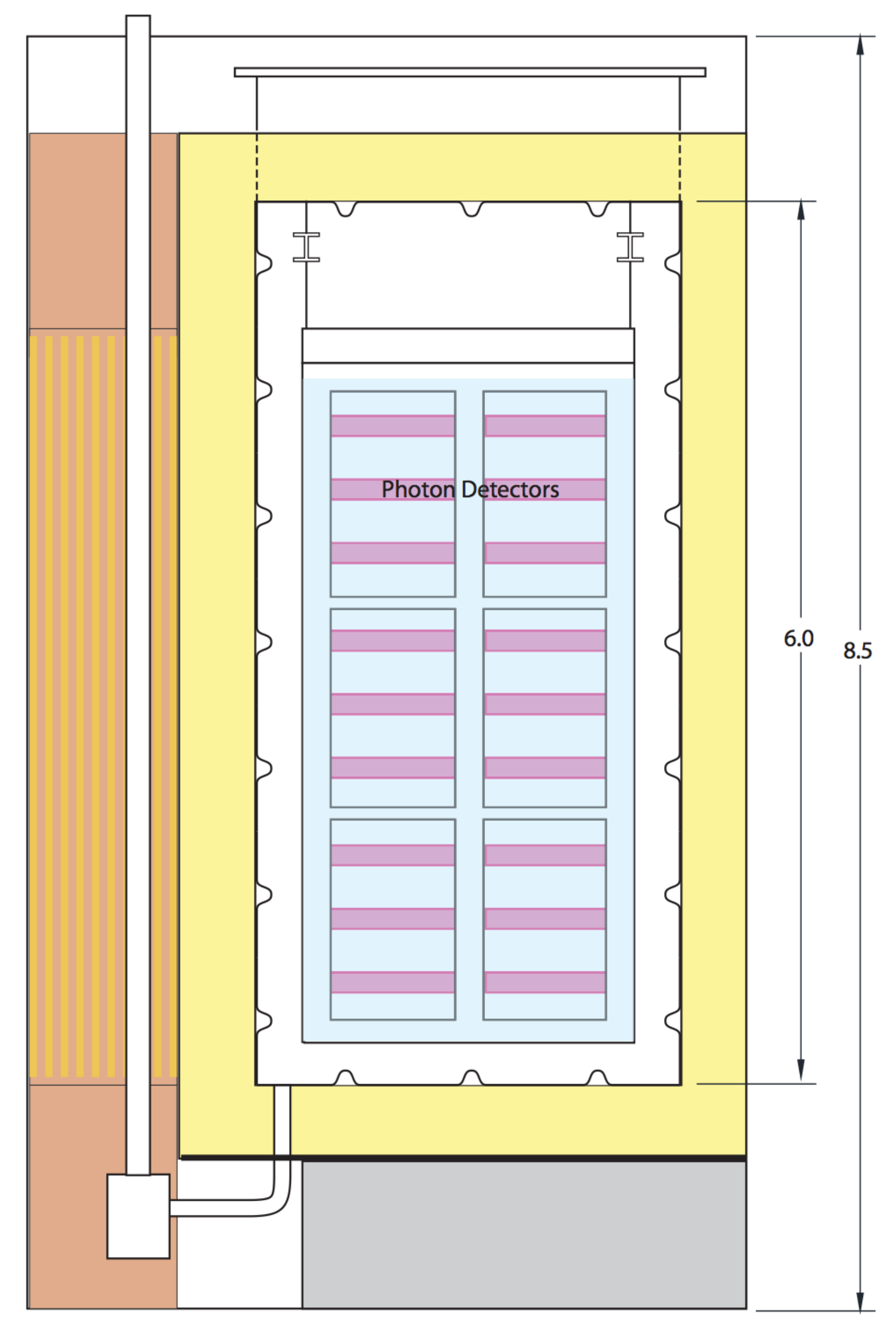}
\caption{\small Side view schematic drawing of the \larnd detector concept.  A membrane cryostat construction is built to fill the existing enclosure.  0.5~m of foam insulation surrounds the corrugated steel membrane.  The cryostat is raised by ~1m using shielding blocks to allow the LAr pump to be installed under the cryostat.  The TPC is comprised of anode plane assemblies with embedded photon detectors.}
\label{fig:larnd_det2}
\end{figure}

Figures \ref{fig:larnd_det1} and \ref{fig:larnd_det2} present a schematic design for the \larnd detector.  A corrugated steel membrane cryostat is built up against three of the outer concrete walls of the enclosure.  Downstream of the TPC, a 1.0~m space is left for the installation of a muon detector.  Figure \ref{fig:35ton} shows a photo of the 35 ton membrane cryostat prototype that has been constructed at Fermilab.  This successful test has verified this technology for use in detectors of this scale and is the intended technology for use in LBNE. 

The details of the muon detector are still under development, but the \SB experiment included a downstream muon range detector (MRD) constructed of 2 inch thick steel plates interleaved with segmented scintillator planes.  The total thickness of the \SB MRD was 69~cm.  The detector used in \larnd will have the additional requirement of being structurally capable of supporting the hydrostatic pressure of the downstream wall of the membrane cryostat.  In the conceptual layout shown in Fig. \ref{fig:larnd_det1} a 1~m $\times$ 1~m well is allocated next to the muon detector to house an argon pump needed near the bottom of the cryostat floor and to allow it to be serviced.  The cryostat is raised by ~1m using shielding blocks to allow the pump to be installed under the cryostat as shown in Fig. \ref{fig:larnd_det2}.   The total active volume of the configuration shown is $2.3~\m \times (2.7*2)~\m \times 4.5~\m \sim 55~\m^3$, or $\sim$77 tons of liquid argon.


The conceptual design for the TPC and light collection systems follow the LBNE design of wrapped Anode Plane Assemblies (APAs) with embedded photon detectors (PDs).  The APAs can be smaller versions of the LBNE APA design (2.3 m vs. 2.5 m in width, 4.5 m vs. 7 m in height). Each PD assembly is comprised of roughly 1 m long, 10 cm wide plastic bars with SiPMs at one end. The number of PD units required in each APA will be determined by their light collection efficiency.  Shown in the schematics are 18 PD assembles.   With the exception of the frame, all parts of the \larnd APAs could be identical to that of the LBNE design.  


\begin{figure}[h!]
\centering
\mbox{ \includegraphics[width=0.76\textwidth]{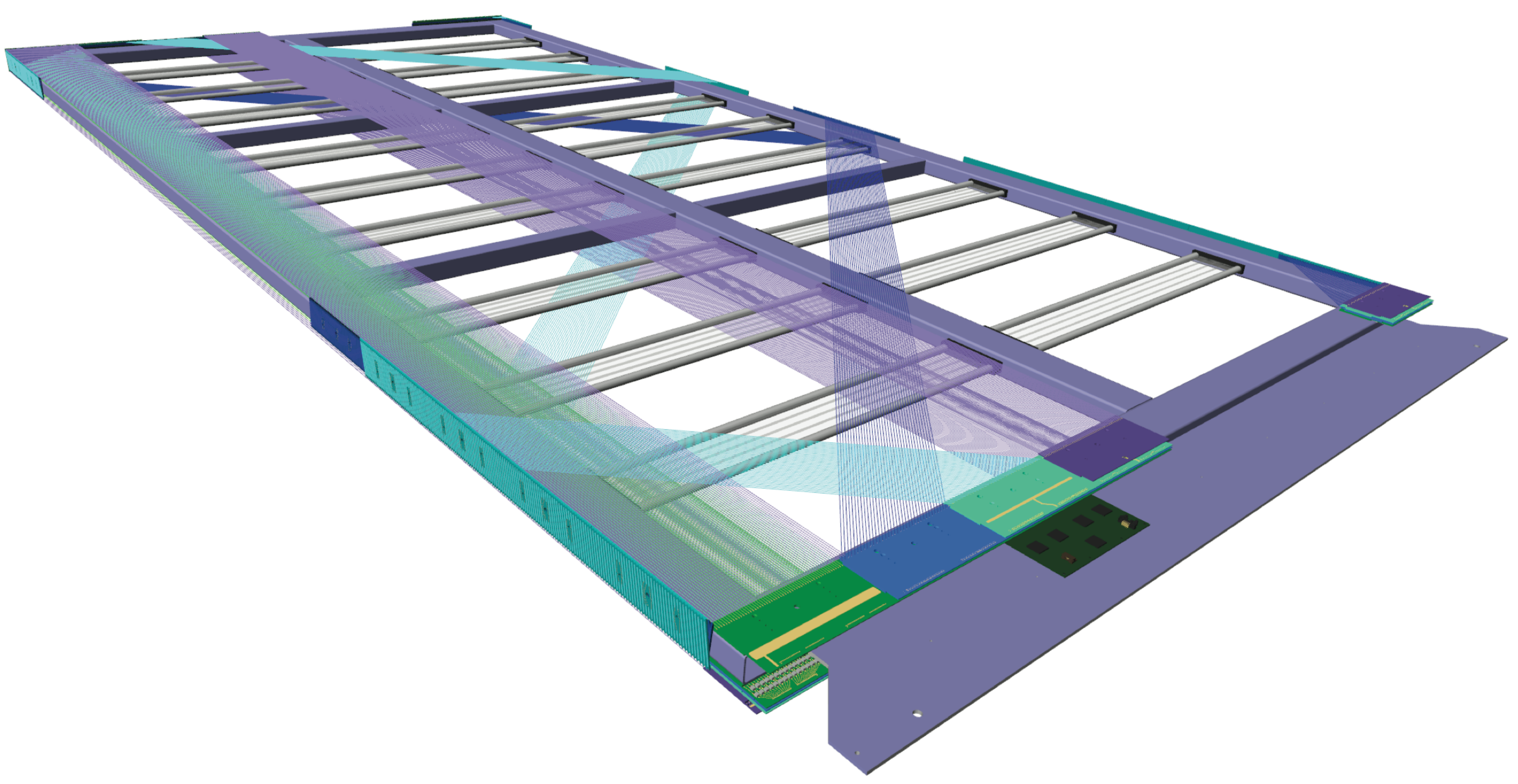} }
\caption{\small This rendering shows a partially assembled APA for the \larnd TPC with 18 photon detector modules. The APA is a smaller version of the LBNE APA design: 2.3m(W) $\times$ 4.5m(H).}
\label{fig:apa}
\end{figure}


Depending on the wire pitch used, the number of channels would be in the range 7,000-10,000.  The large number of readout channels required to instrument the \larnd TPC motivates the use of CMOS ASICs for the electronics.  Both analog FE ASIC and ADC ASIC, to a large extent, have already been developed for LBNE, and analog FE ASIC is being used in MicroBooNE.  The entire front end electronics chain would be immersed in the LAr and operate at 89 K to achieve the optimum signal to noise ratio.

The basic conceptual design briefly outlined here demonstrates that a detector capable of fitting in the existing enclosure along the BNB would be capable of achieving valuable physics goals while implementing key techniques planned for the LBNE far detectors.     

\section{Phase II: Three \lartpc Detector Configuration}
\label{sec:PhaseII}

\larnd presents a compelling physics program and an excellent opportunity to continue the development of \lartpc technology in a running neutrino experiment.   The near detector can be constructed quickly and at modest cost, and, in combination with \uboone, will be able to make important statements regarding existing anomalies seen in neutrinos and record more than 600k neutrino interactions per year for studying neutrino-argon interactions. 

\larnd can also be thought of as the next phase in the development of a world-class program of short-baseline accelerator-based neutrino physics at Fermilab.  The addition of a large (kton-scale) detector at a longer baseline ($\sim$700~m) would address oscillations in antineutrinos and make precision measurements of sterile neutrino oscillations if they are discovered.   This three detector configuration, depicted in Fig. \ref{fig:baseline}, is extremely powerful for addressing this physics with \larnd measuring the unoscillated neutrino fluxes to constrain systematic errors, a large-scale far detector measuring an oscillation signal with excellent significance and high statistics, and a signal in \uboone in the middle to confirm the interpretation as new physics.  

\begin{figure}[h]
\centering
\mbox{ \includegraphics[width=0.42\textwidth]{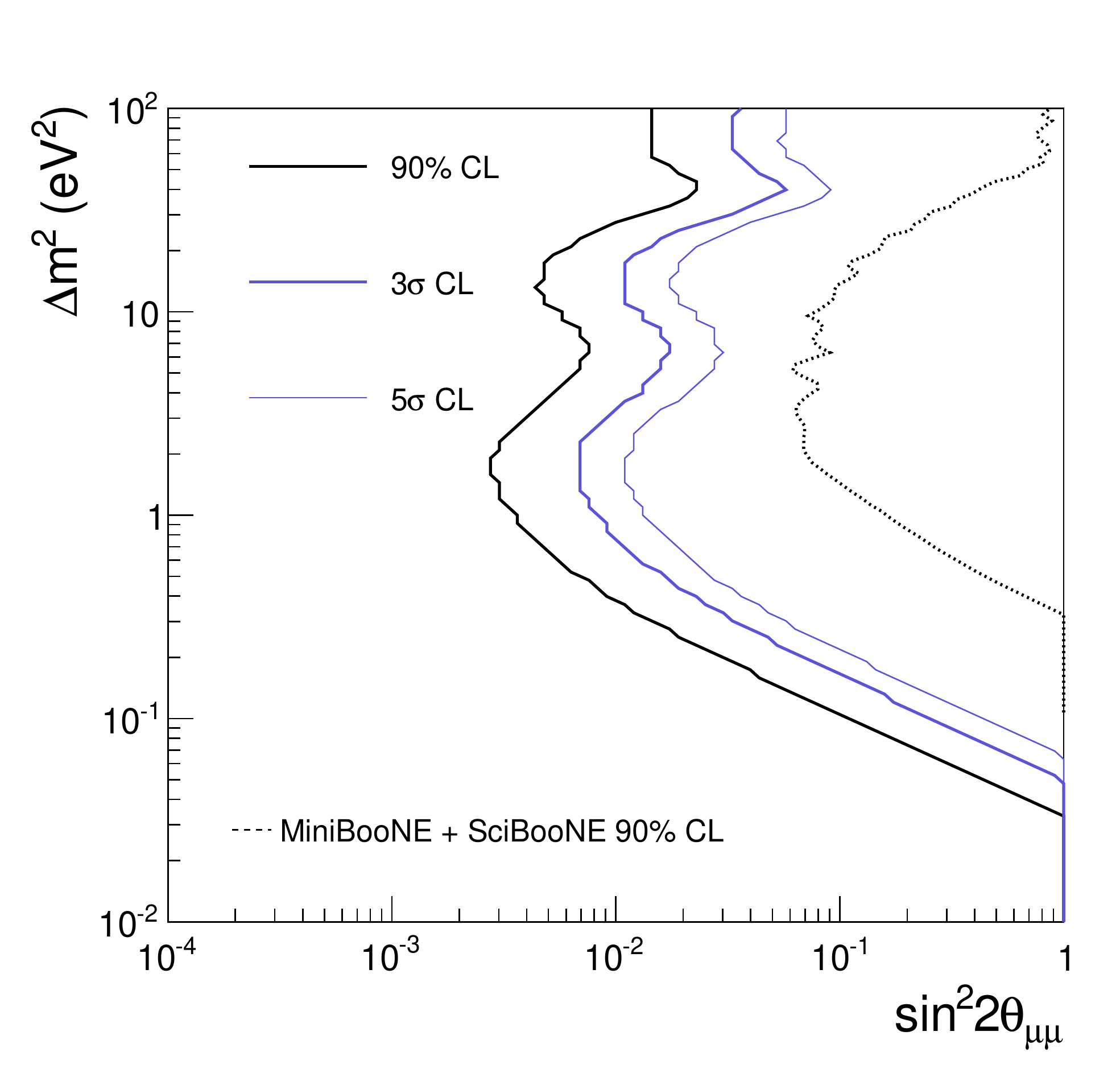}
\includegraphics[width=0.59\textwidth]{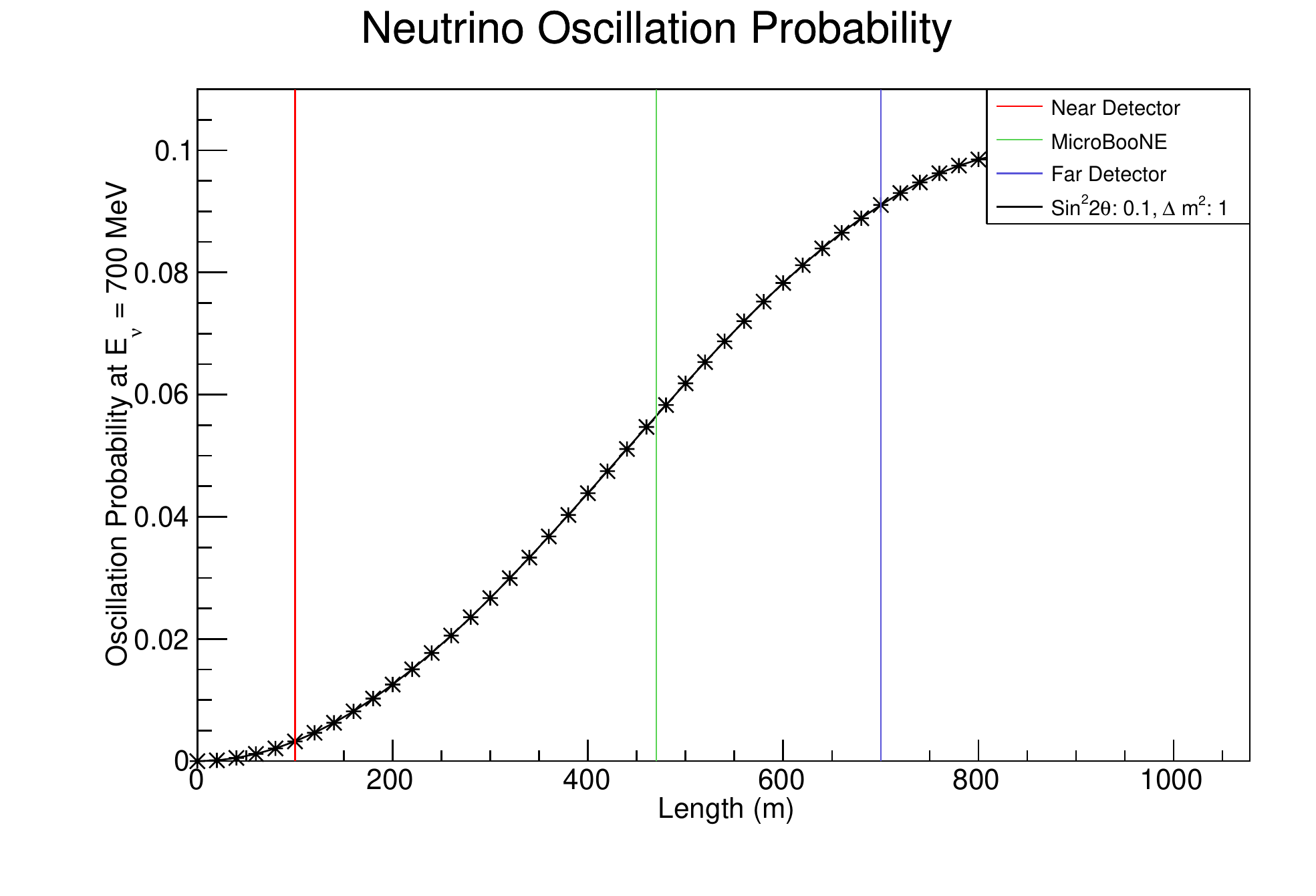}}
\caption{ \small Sensitivity to \numu disappearance with the full \larone experiment, a program of three \lartpc detectors on the Booster Neutrino Beamline at Fermilab (left).  \numu disappearance probability at $\Enu = 700~\MeV$ as a function of distance in a sterile neutrino model with $\dmsq = 1~\eVsq$ and $\sinth_{\mu\mu} = 0.1$ (right).  The vertical colored lines indicate the proposed locations of \larnd, \uboone and \larfd.}
\label{fig:numudis_phase2}
\end{figure}

The LAr1 program will definitively address the short baseline anomalies observed by \MB in both neutrino and antineutrino mode.  Detailed studies combined with complete conceptual design of the detector are underway and will be described in a full proposal to be submitted later this year.  As an example of the reach of the detector, Fig. \ref{fig:numudis_phase2} provides an example of the strength of the full \larone program.  Shown is the sensitivity to \numu disappearance through charged-current interactions using \larnd, \uboone and a 1 kton fiducial volume detector at 700~m.  This is the shape-only sensitivity to be compared to the lower left plot in Fig. \ref{fig:numudis_phase1}.  This configuration enables 5$\sigma$ sensitivity down to a few percent \numu disappearance in the $\dmsq = 1~\eVsq$ region.  

The right panel in Fig. \ref{fig:numudis_phase2} depicts the evolution of the oscillation probability with distance for oscillations in a 3+1 model occurring at $\dmsq_{41} = 1~\eVsq$.  The probability is shown for the peak neutrino energy, 700~MeV.  The vertical lines indicate the locations of the detectors and illustrate the ability to measure the changing oscillation with three detectors.  This will provide a powerful method for the interpretation of any observed signals within an oscillation model.  Two additional things to note.  First, for \dmsq less than a few \eVsq, the location of \larnd at 100~m is ideal for sampling the neutrino flux before the onset of any significant oscillation.  This, and the factor 15 higher event rate compared to \uboone, makes \larnd a critical aspect of the strength of the overall program.   Second, significant sensitivity is regained for large values of \dmsq ($\geq 10~\eVsq$) in this shape-only analysis relative to the 2-detector configuration.  Sensitivities for electron neutrino appearance in neutrino and antineutrino mode will be presented in the full proposal.   




\section{Conclusions}
\label{sec:conclusion}

The hints for new physics at short baselines are being explored by new experiments in the planning stages and near running worldwide.  Whether or not we find new physics, the anomalies must be resolved definitively.  Fermilab will have the first opportunity to address these with MicroBooNE.   Beyond this, the LAr1 program will confirm or rule out the anomalies in both neutrino and anti-neutrino mode to 5 sigma.  LAr1-ND, the first phase in this full program beyond MicroBooNE, will both interpret and characterize a MicroBooNE signal, be it electrons or photons, and serve as the first step for measurements in anti-neutrino mode as a near detector for the LAr1 far detector.  

This white paper describes LAr1-ND and the compelling physics it brings first in Phase 1 and next towards the full LAr1 program.  In addition, LAr1-ND serves as a key step in the development toward large-scale LArTPC detectors.   Its development goals will encompass testing existing and possibly innovative designs for LBNE while at the same time providing a training ground for teams working towards LBNE combining timely neutrino physics with experience in detector development.  

This white paper is being developed into a proposal to be submitted to the Fermilab Director in time for the fall 2013 Fermilab PAC.

\clearpage

\end{document}